\documentclass[aps,prb,reprint,superscriptaddress,amsmath]{revtex4-2}
\usepackage{graphicx}
\usepackage{refcount}
\usepackage{fmtcount}
\usepackage{amssymb}
\usepackage{dsfont}
\usepackage{hyperref}
\usepackage[usenames,dvipsnames]{color}
\usepackage[normalem]{ulem}
\usepackage{tikz}
\usepackage{bm}
\usepackage{pdfpages}

\makeatletter
\AtBeginDocument{\let\LS@rot\@undefined}
\makeatother

\hypersetup{pdfstartview=FitH,pdfpagemode=UseOutlines,colorlinks,
citecolor=blue,linkcolor=blue,urlcolor=blue}

\graphicspath{{.}{./Figures/}}

\renewcommand{\a}{\alpha}
\renewcommand{\b}{\beta}

\newcommand{\s}{\sigma}
\renewcommand{\t}{\theta}
\newcommand{\dg}{\dagger}

\newcommand{\llangle}{\langle\!\langle}
\newcommand{\rrangle}{\rangle\!\rangle}

\definecolor{blue(munsell)}{rgb}{0.0, 0.5, 0.69}

\newcommand{\add}[1]{ { \color{blue}  #1 }}

\definecolor{aaron}{rgb}{0.6, 0.6, 0.8}

\begin{document}

\title{Phase diagram of a bilinear-biquadratic spin-1 model on the triangular lattice
from density matrix renormalization group simulations}

\author{Aaron Szasz}
	\email[]{aszasz@perimeterinstitute.ca}
	\affiliation{Perimeter Institute for Theoretical Physics, Waterloo, Ontario, N2L 2Y5, Canada}
\author{Chong Wang}
	\affiliation{Perimeter Institute for Theoretical Physics, Waterloo, Ontario, N2L 2Y5, Canada}
\author{Yin-Chen He}
	\affiliation{Perimeter Institute for Theoretical Physics, Waterloo, Ontario, N2L 2Y5, Canada}

\date{\today}

\begin{abstract}
We investigate a highly frustrated spin-1 model on the triangular lattice, with nearest- and next-nearest-neighbor antiferromagnetic $S\kern -0.03em \cdot\kern -0.03em S$  interactions and nearest-neighbor $(S\kern -0.03em \cdot\kern -0.03em S)^2$ interactions. Using the density matrix renormalization group (DMRG) technique, we find three magnetically ordered phases, namely 120$^\circ$ spiral order, stripe order, and tetrahedral order, as well as two spin nematic phases: ferroquadrupolar and antiferroquadrupolar.  While our data could be consistent with a spin liquid phase between the 120$^\circ$ spiral and antiferroquadrupolar orders, the more likely scenario is a direct continuous transition between these two orders.
\end{abstract}

\maketitle


\section{Introduction:}
Geometric frustration of antiferromagnetism can lead to novel phases of matter such as spin liquids, which preserve all symmetries of the system down to zero temperature and feature gapless or topological quasiparticle excitations~\cite{Balents2010,Savary2017,Zhou2017}.  Such a spin liquid state was first predicted by Anderson in 1973 for the antiferromagnetic nearest-neighbor spin-1/2 Heisenberg model on the triangular lattice~\cite{Anderson1973}.  Although that particular model was later shown to realize a three-sublattice magnetic order in the ground state~\cite{Huse1988,White2007}, in 2003 spin liquid-like behavior was indeed observed in a material approximately described by weakly coupled triangular lattice layers, $\kappa$-(BEDT-TTF)$_2$Cu$_2$(CN)$_3$~\cite{Shimizu2003}, which was found to have no magnetic ordering down to low temperature.  Subsequent experiments demonstrated further properties suggestive of a (possibly topological) spin liquid, including specific heat consistent with gapless low-energy excitations~\cite{Yamashita2008} and thermal conductivity consistent with a gapped bulk~\cite{Yamashita2008b}.

As $\kappa$-(BEDT-TTF)$_2$Cu$_2$(CN)$_3$ and other spin liquid-candidate materials such as EtMe$_3$Sb[Pd(dmit)$_2$]$_2$~\cite{Itou2008,Itou2009,Yamashita2010,Itou2010,Ni2019,BourgeoisHope2019,Yamashita2019}, YbMgGaO$_4$~\cite{Li2016,Shen2016,Shen2017}, and herbertsmithite~\cite{Helton2007, Han2012} have one free electron per effective lattice site, theory work has largely focused on the half-filled Hubbard model and its strong-coupling limit, the spin-1/2 Heisenberg model with higher-order terms beyond nearest neighbor.  Recent numerical studies confirm that these models indeed have spin liquid ground states, including chiral spin liquids and Dirac spin liquids on the triangular~\cite{Iqbal2016,Hu2019b,Szasz2020, Szasz2021, Cookmeyer2021, Chen2021, Zhou2022} and kagome lattices~\cite{Ran2007,Iqbal2013, He_chiral, Gong_chiral, He2017}.

\begin{figure}
\includegraphics[width = 0.48\textwidth]{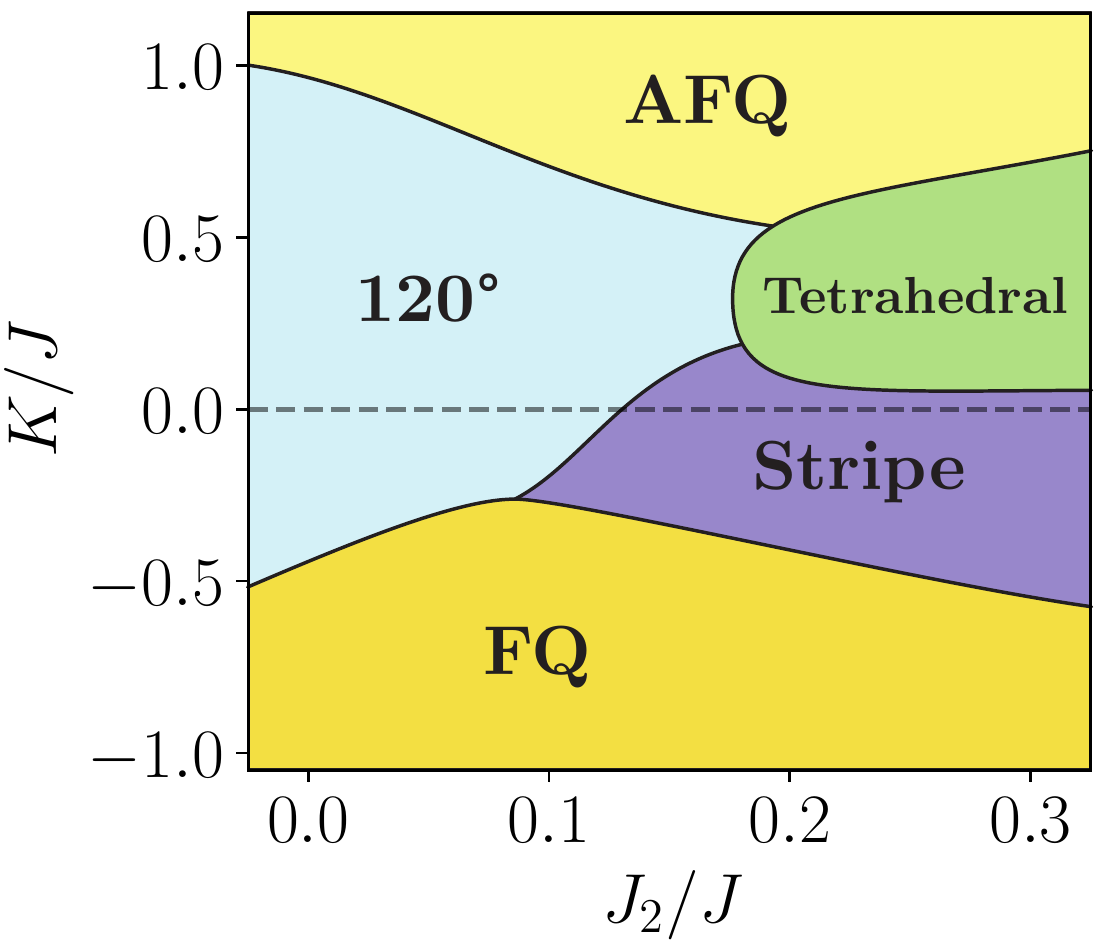}
\caption{(Color online) Phase diagram of spin-1s on the triangular lattice with three competing interactions, based on simulations on a circumference-six cylinder with YC boundary conditions.  Axes are strength of next-nearest-neighbor interactions, $J_2$, and nearest-neighbor biquadratic interactions, $K$, relative to antiferromagnetic nearest-neighbor interactions, $J$.  There are three magnetic orders (120$^\circ$, stripe, and tetrahedral) and two nematic orders, ferroquadrupolar (FQ) and antiferroquadrupolar (AFQ).  We do not find a disordered phase. 
\label{fig:PD}}
\end{figure}

There are also materials realizing higher-spin frustrated antiferromagnets.  In 2005, Nakatsuji \textit{et al.} discovered a possible spin-liquid candidate material that can be approximately described by a spin-1 triangular lattice, NiGa$_2$S$_4$, which also shows a lack of magnetic order down to low temperature~\cite{Nakatsuji2005, Nambu2006, Nakatsuji2007, Takeya2008, MacLaughlin2008,Nakatsuji2010,MacLaughlin2010}.  First-principles electronic structure calculations and other methods have suggested that NiGa$_2$S$_4$ has strong third-neighbor Heisenberg (bilinear) interactions, as well as significant nearest-neighbor and possibly second-neighbor interactions~\cite{Mazin2007, Takubo2007, Takubo2009, Takenaka2014, Takenaka2016}.  Furthermore, the most general two-body interactions for spin-1 systems include $({\bf S}\cdot{\bf S})^2$ (biquadratic) interactions~\cite{Chubukov1990}, which are believed to be significant in a variety of materials~\cite{Kartsev2020}; following an initial proposal by Tsunetsugu and Arikawa~\cite{Tsunetsugu2006}, nearest-neighbor biquadratic interactions are typically included in phenomenological models of NiGa$_2$S$_4$.  

The spin-1 triangular lattice model with nearest-neighbor bilinear and biquadratic terms has been studied extensively since the mid 2000s, and has been shown to host four phases: the same three-sublattice magnetic order found in the spin-1/2 triangular lattice, ferromagnetic order, and two types of nematic order~\cite{Lauchli2006,Moreno-Cardoner2014,Niesen2017,Niesen2018}.  These nematic states have been proposed to explain the behavior of NiGa$_2$S$_4$~\cite{Tsunetsugu2006,Bhattacharjee2006,Tsunetsugu2007}, but the matter remains unresolved.

One of our goals in the present work is to generalize these spin-1 triangular lattice studies by incorporating longer-ranged interactions.  The specific model we study is presented in Eq.~\eqref{eq:H} below; in summary, the model has antiferromagnetic Heisenberg (bilinear) interactions between nearest and next-nearest neighbors, and biquadratic interactions (of both signs) between nearest neighbors.  While we consider a second-neighbor interaction rather than the (in principle more relevant to NiGa$_2$S$_4$) third-neighbor interaction, our results still shed light on how the nearest-neighbor-only phase diagram changes when additional terms are included.  Additionally, second-neighbor interactions may be relevant to other spin-1 triangular lattice systems, including the spin liquid candidate Ba$_3$NiSb$_2$O$_9$~\cite{Doi2004, Cheng2011, Ono2011, Shirata2011, Quilliam2016, Fak2017} and materials with magnetic ground states such as Ba$_2$La$_2$NiTe$_2$O$_{12}$~\cite{Saito2019}, Na$_2$BaNi(PO$_4$)$_2$~\cite{Li2021}, and FeI$_2$~\cite{Bai2021}.

We have two key motivations for including specifically the second-neighbor interaction.  First, a large-$S$ expansion of precisely this model shows the possible appearance of a quantum disordered phase as $S$ becomes small~\cite{Yu2020}.  Second, we aimed to study a model that could feature a Stiefel liquid; Stiefel liquids are a recently predicted novel class of disordered quantum phases generalizing the Dirac spin liquid and deconfined quantum critical point~\cite{Zou2021,Ye2021}.  The simplest never-yet-observed example of a Stiefel liquid should occur in proximity to the non-coplanar tetrahedral magnetic order~\cite{Zou2021}, while another Stiefel liquid could arise near a nematic phase~\cite{Ye2021}.  The spin-1 model we consider is expected, based on the large-$S$ predictions and past results on the nearest-neighbor model, to host both ordered phases, and hence could plausibly realize these Stiefel liquids.  

We study the extended bilinear-biquadratic model using density matrix renormalization group (DMRG)~\cite{White1992, White1993, Schollwock2011, TenPy2} simulations on infinitely-long finite circumference cylinders; DMRG finds the lowest energy state within the variational class of matrix product states (MPS).  We find precisely all five expected phases, namely the three magnetic orders (120$^\circ$, stripe, tetrahedral) from the classical limit and the two nematic orders (ferroquadrupolar, antiferroquadrupolar) found in the limit of no second-neighbor interaction.  Our phase diagram is shown in Figure~\ref{fig:PD}.  Our results are plausibly consistent with a disordered phase, which could be a spin liquid, between the 120$^\circ$ and antiferroquadrupolar phases, but there is no compelling evidence favoring this interpretation over a direct continuous phase transition.  Our results do suggest that there is no first-order transition between these phases, in contrast to the conclusions of some previous works~\cite{Niesen2017,Niesen2018}.

The remainder of the paper is organized as follows.  In Section \ref{sec:model}, we introduce our model and comment on our DMRG implementation.  We also provide a review of both spin-quadrupole order and past results from the literature.   In Section \ref{sec:phase_diagram} we present detailed results from our simulations, including key data such as structure factors and correlation lengths that we use to identify the phases and locate their boundaries; we explain why we interpret our data as indicating the absence of a disordered phase.  Finally, in Section \ref{sec:discussion}, we summarize our findings and discuss possible modifications to the model that could lead to spin liquid ground states.


\section{The Model:\label{sec:model}}

We study spin-1 degrees of freedom on the triangular lattice, with the Hamiltonian
\begin{equation}
H = \sum_{\langle i j\rangle}\left(J\, {\bf S}\cdot {\bf S} + K({\bf S}\cdot {\bf S})^2\right) + J_2\sum_{\llangle i j \rrangle} {\bf S}\cdot {\bf S},\label{eq:H}
\end{equation}
where $\langle i j \rangle$ denotes pairs of nearest neighbor sites and $\llangle i j \rrangle$ denotes next-nearest neighbors; ${\bf S}$ on each site is the usual spin-1 operator.  We consider only the case of $J$, $J_2>0$, but allow $K$ to have both signs. Throughout the paper, we refer to the ${\bf S}\cdot {\bf S}$ and $({\bf S}\cdot {\bf S})^2$ interactions as bilinear and biquadratic, respectively.  

\begin{figure}
\centering
\includegraphics[width=0.48\textwidth]{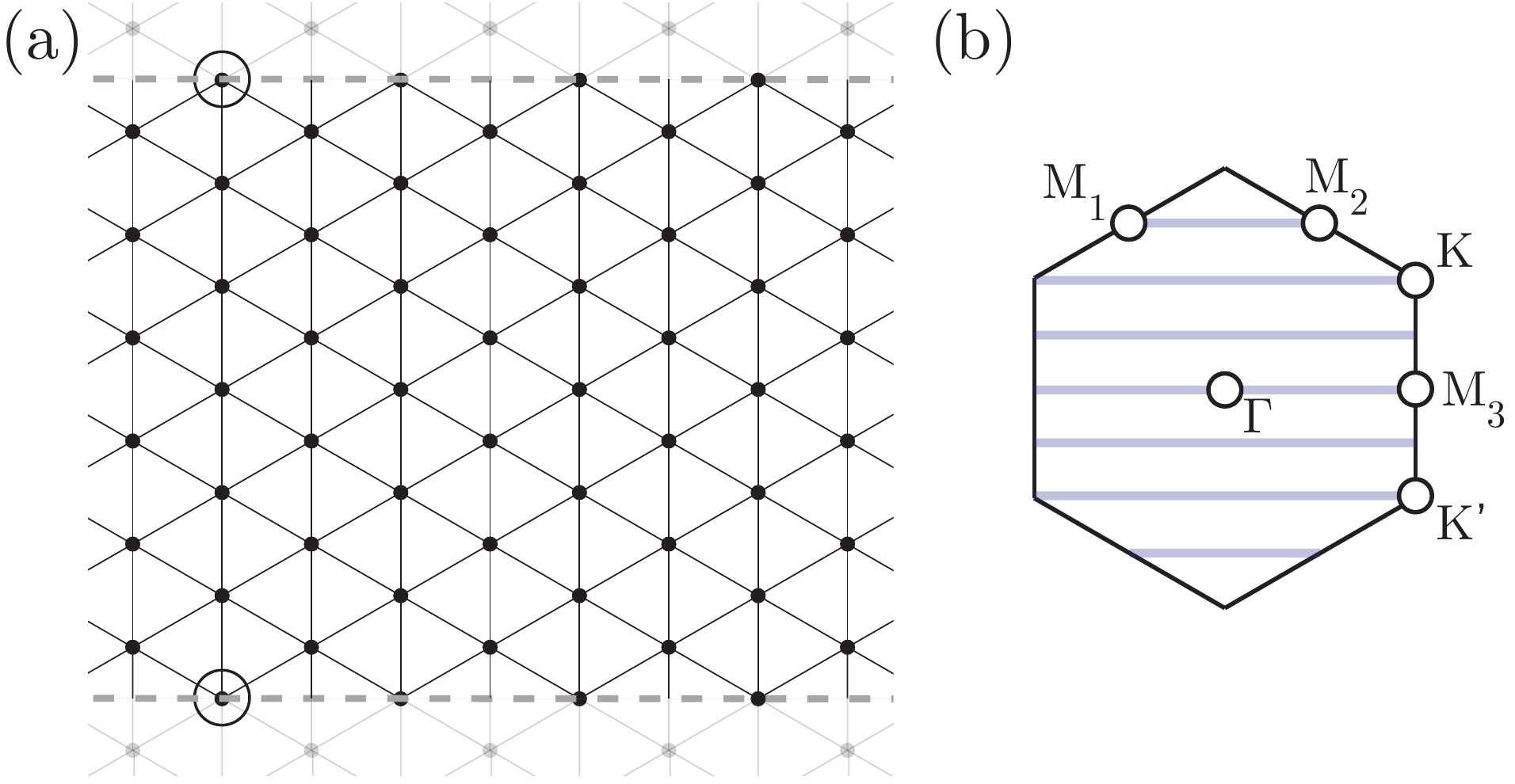}
\caption{(a) The YC6 triangular lattice.  There are periodic boundary conditions in the direction along the vertical bond, so that the two dashed lines are identified together to form a cylinder.  The two circled points are equivalent. (b) The corresponding Brillouin zone.  The finite circumference cylinder restricts $k_y$ to lie on the horizontal cuts shown.  For the YC6 cylinder in particular, these cuts contain all the high-symmetry points indicated in the figure: $\Gamma$ at the zone center, $K$ and $K'$ at the corners, and the three $M$ points at the centers of the edges.
\label{fig:lattice}}
\end{figure}

We study this model using DMRG simulations on infinite cylinders with a finite circumference of six sites, using YC boundary conditions~\cite{Szasz2020} as shown in Figure~\ref{fig:lattice}(a).  We improve the efficiency of our simulation by using delinearisation~\cite{Hubig2017} to perform lossless compression of the matrix product operator representation of $H$.  In momentum space, the finite cylinder circumference discretizes the allowed momentum in the direction around the cylinder.  In Figure~\ref{fig:lattice}(b), we show the allowed momenta in the Brillouin zone for the cylinder we consider.  Importantly, the allowed momentum cuts for the YC6 cylinder include all high-symmetry points in the Brillouin zone, namely the M, K, and $\Gamma$ points as labeled in the figure.

Before proceeding to discuss our findings, we review the expected behavior of the model, in limiting cases and based on results from the literature.

\subsection{Limiting cases and quadrupolar order}

We consider the following three limits of the model: $J>0$ only, $K>0$ only, and $K<0$ only.  We also discuss the parameter point $J=K$, $J_2=0$, where there is an emergent larger global symmetry group. 

For the pure nearest-neighbor Heisenberg antiferromagnet, with $K=J_2=0$, the ground state has $120^\circ$ three-sublattice magnetic order.  This is the classical (large-$S$) ground state, and has also been established by various numerical methods including exact diagonalization~\cite{Lauchli2006}, cluster mean-field theory~\cite{Moreno-Cardoner2014}, and tensor network simulations~\cite{Niesen2017,Niesen2018}, as the ground state for spin 1.  In momentum space, the spin structure factor
\begin{equation}
S({\bf q}) = \frac{1}{N}\sum_{ij}e^{i{\bf q}\cdot {\bf r}_{ij}}\left\langle {\bf S}_i{\bf S}_j\right\rangle\label{eq:def_Sq}
\end{equation}
has peaks at the K and K' points at the corners of the hexagonal Brillouin zone. 

More interesting is the purely biquadratic Hamiltonian, with $J=J_2=0$, which gives rise to spin-nematic order~\cite{Blume1969, Chen1971, Matveev1974, Andreev1984, Papanicolaou1988}.  Like a magnetically-ordered state, a spin-nematic state breaks spin rotation symmetry; unlike a magnetic state, it preserves time-reversal.  This combination is possible if the spin dipole moment is zero, but higher-order multipoles are nonzero.  Quadrupole order in particular is very natural for spin 1, as we now explain.

The key insight is that each spin-1 Hilbert space has an orthonormal basis of spin-quadrupole states, namely, the 0-eigenvectors of $\hat{n}\cdot{\bf S}$ along any three orthogonal directions.  For example, the 0-eigenvectors of $S_z$, $S_x$, and $S_z$ are, when written in the $z$-basis, $|z\rangle\equiv -i(0,1,0)^T$, $|x\rangle\equiv(i,0,-i)^T/\sqrt{2}$, and $|y\rangle \equiv (1,0,1)^\text{T}/\sqrt{2}$; the somewhat strange coefficients are chosen for convenience, following ref.~\cite{Lauchli2006}.

\begin{figure}
\centering
\includegraphics[width=0.48\textwidth]{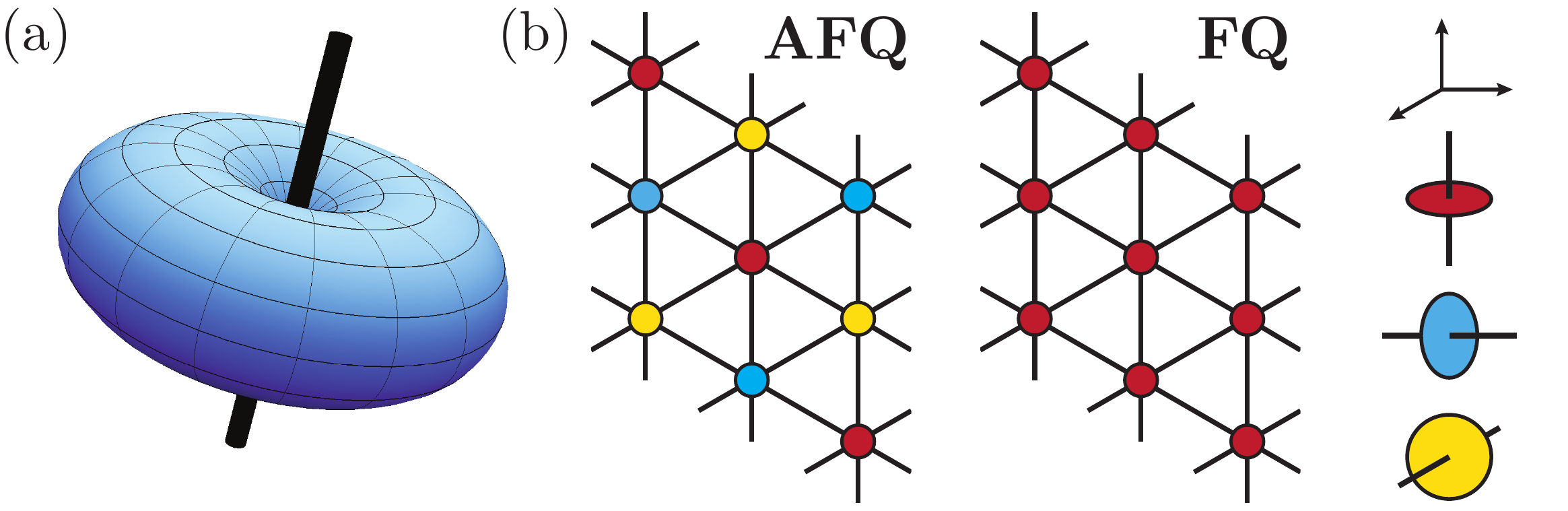}
\caption{(Color online) (a) A visualization of single-site spin-quadrupole order.  The distance of the surface from the origin in each direction $\hat{n}$ shows $\langle (\hat{n}\cdot{\bf S})^2 \rangle$.  Spin fluctuations are small along the ``director'' and large in the two perpendicular directions.  (b) AFQ and FQ order on the triangular lattice.  The cartoon picture of AFQ order is a three-sublattice product state of single-site quadrupoles; the director is the same for all sites within a sublattice, and the directors on the three sublattices lie along three orthogonal directions.  The quadrupole structure factor has peaks at the $K$ points. The cartoon picture of the FQ order is a product state of single-site quadrupoles with parallel directors on all sites.  The quadrupole structure factor has a peak at the $\Gamma$ point.
\label{fig:nematic}}
\end{figure}

Each of these 0-eigenvectors is a single-site (quadrupolar) nematic state.  Consider, for example, $|z\rangle$.  This state has $\langle z|\hat{n}\cdot{\bf S}|z\rangle=0$ for all $\hat{n}$, so time-reversal symmetry is not broken; however, $\langle z|(\hat{n}\cdot{\bf S})^2|z\rangle = \sin^2(\theta)$ where $\t$ is the angle between $\hat{z}$ and $\hat{n}$.  As illustrated in Figure~\ref{fig:nematic}(a), spin-rotation symmetry is broken to the symmetry of an ellipsoid, which has a quadrupole moment but no dipole moment.  More generally, the state $|n\rangle = \hat{n}\cdot(|x\rangle,|y\rangle,|z\rangle)$, for any real unit vector $\hat{n}$, is the 0-eigenstate of $\hat{n}\cdot{\bf S}$, and hence is likewise a spin-quadrupole.  The direction with no spin fluctuations, $\pm\hat{n}$, is called the ``director'' of the nematic state.

To more precisely characterize quadrupolar nematic order, we use the spin-quadrupole operator $\mathbf{Q}$ given by
\begin{equation}
\frac{Q_{\a\b}}{\sqrt{2}}=\frac{S_\a S_\b + S_\b S_\a}{2} - \frac{S(S+1)}{3}\delta_{\a\b}\text{Id}\label{eq:def_Q},
\end{equation}
for $\a$,$\b\in \{x,y,z\}$, a traceless symmetric \add{rank-2} tensor.~\footnote{The factor of $\sqrt{2}$ on the LHS is included for consistency with ${\bf Q}_v$, Eq.~\eqref{eq:def_Q_v2}.  Note that some past works, such as reference~\cite{Kaul2012}, define ${\bf Q}$ without this $\sqrt{2}$.}  The symmetry and trace conditions imply there are only five independent components, so ${\bf Q}$ can equivalently be written as the vector: 
\begin{equation}
\mathbf{Q}_v=\left(\begin{array}{c}
S_x^2-S_y^2\\ (2Sz^2-S_x^2-S_y^2)/\sqrt{3}\\ \{S_x, S_y\}\\ \{S_y, S_z\}\\ \{S_x, S_z\}\label{eq:def_Q_v2}
\end{array}\right)
\end{equation}
with $\{\,,\,\}$ the anticommutator.  The square of the quadrupole operator is $\text{Tr}[{\bf Q}{\bf Q}]={\bf Q}_v\cdot{\bf Q}_v = (4/3)\text{Id}$.  (Note that the trace is taken over the components of the operator-valued matrix ${\bf Q}^2$, and not over the spin operators.)  It follows that no component of ${\bf Q}_v$ can have an eigenvalue with magnitude greater than $\sqrt{4/3}$; the basis state $|z\rangle$ has $\langle{\bf Q}_v\rangle=(0,-\sqrt{4/3},0,0,0)$, so it indeed realizes the maximum possible single-site quadrupole moment. 

Let us now consider the interaction $({\bf S}\cdot{\bf S})^2$.  We can write this as~\cite{Kaul2012}
\begin{equation}
({\bf S}\cdot{\bf S})^2 = 1+3|S\rangle\langle S|,\,\,\,\,\,\,|S\rangle = \frac{|zz\rangle + |xx\rangle + |yy\rangle}{\sqrt 3}.
\end{equation}
The projector $P=|S\rangle\langle S|$ is positive semi-definite, so when $K>0$ the ground state will be in the space projected out by $P$ on every pair of sites.  Thus any product state where each site is in the state $|x\rangle$, $|y\rangle$, or $|z\rangle$, and where no two nearest neighbors are in the same state, will be a ground state.  We produce such a state on the triangular lattice by identifying three sublattices and placing $|x\rangle$ on each site of one sublattice, and likewise for $|y\rangle$ and $|z\rangle$ on the other two sublattices, as shown in Figure~\ref{fig:nematic}(b).  The ordering of this state is referred to as antiferroquadrupolar (AFQ).  In momentum space, it is characterized by the quadrupolar structure factor,
\begin{equation}
Q({\bf q}) = \frac{1}{N}\sum_{ij}e^{i{\bf q}\cdot {\bf r}_{ij}}\left\langle \text{Tr}[{\bf Q}_i{\bf Q}_j]\right\rangle\label{eq:def_Qq},
\end{equation}
which has peaks at the $K$ points.

However, the ground state space also contains ferromagnetic (FM) states.  For example, the state $|1\rangle_z\equiv(|y\rangle-i|x\rangle)/\sqrt{2}$, with $S_z=1$, has $P\left(|1\rangle_z\otimes |1\rangle_z\right)=0$, so the fully magnetized state with $|1\rangle_z$ on every site is also a ground state.  Thus the pure $K>0$ model has an exact degeneracy between FM and AFQ order.  Infinitesimal nearest-neighbor interactions break the degeneracy: $J>0$ gives AFQ order, while $J<0$ gives FM order, and the transition is first-order.  We consider only $J>0$, so we expect AFQ order in the large-$K$ limit.

When $K<0$, a classical or large-$S$ ground state would be any in which each spin points either parallel or anti-parallel to its neighbors, behavior which we call ferro-nematic or ferroquadrupolar (FQ).  In the case of spin-1, a product state will no longer be a ground state, but sign-free quantum Monte Carlo (QMC) calculations provide clear evidence that the ground state remains FQ~\cite{Kaul2012}.  The prototypical FQ state for spin-1 is the product state with $|z\rangle$ on every site, which has a quadrupole structure factor with a peak at the $\Gamma$ point, ${\bf k}={\bf 0}$; this state is shown in Figure~\ref{fig:nematic}(b).  From the QMC results, the FQ ground state in the pure $K<0$ model is in the same phase as this product state, but with magnitude of the quadupolar structure factor reduced by about 50\%.

There is one more important parameter point to consider, namely $J_2=0$, $J=K$.  To demonstrate the significance of this parameter point, we rewrite the biquadratic interaction as
\begin{equation}
\left({\bf S}_i\cdot{\bf S}_j\right)^2=\frac{{\bf Q}_{vi}\cdot{\bf Q}_{vj}}{2} - \frac{{\bf S}_i\cdot{\bf S}_j}{2}+\frac{4}{3}\label{eq:biquad_QQ_SS},
\end{equation}
in which case the Hamiltonian becomes, for $J_2=0$,
\begin{equation}
H = \sum_{\langle ij\rangle} \left(J-\frac{K}{2}\right){\bf S}_i\cdot{\bf S}_j + \frac{K}{2}{\bf Q}_{vi}\cdot{\bf Q}_{vj}+\frac{4}{3}\label{eq:QQ_SS_H}.
\end{equation}
Following reference~\cite{Lauchli2006}, when $J=K$ we can further combine the spin and quadrupole operators to get
\begin{equation}
{\bf S}_i\cdot{\bf S}_j + {\bf Q}_{vi}\cdot{\bf Q}_{vj} = 2W_{ij}-(2/3)\text{Id}
\end{equation}
where $W_{ij}$ is the operator that swaps the states of spins $i$ and $j$, $W_{ij} = \sum_{\a\b} |\a\rangle_i|\b\rangle_j\langle\b|_i\langle\a|_j$ where $\a$ and $\b$ index basis states for the two spins.  
The swap operator is invariant under conjugation by $U\otimes U$ for any $U\in SU(3)$:
\begin{equation}
(U\otimes U)W(U^\dg \otimes U^\dg) = 
\begin{tikzpicture}[baseline=(current  bounding  box.center)]
\draw (-0.4, 0.4) -- (-0.2,0.4);
\draw (-0.4, -0.4) -- (-0.2,-0.4);
\draw (-0.2, 0.2) rectangle (0.2, 0.6);
\draw (-0.2, -0.2) rectangle (0.2, -0.6);
\draw node at (0.02,0.39) {$U^{ }$};
\draw node at (0.02,-0.41) {$U^{ }$};
\draw (0.2, 0.4) -- (0.3,0.4);
\draw (0.2, -0.4) -- (0.3,-0.4);

\draw (0.3, -0.4) -- (0.7,0.4);
\draw (0.3, 0.4) -- (0.45,0.1);
\draw (0.55,-0.1) -- (0.7,-0.4);

\draw (0.7, 0.4) -- (0.8,0.4);
\draw (0.7, -0.4) -- (0.8,-0.4);
\draw (0.8, 0.2) rectangle (1.24, 0.6);
\draw (0.8, -0.2) rectangle (1.24, -0.6);
\draw node at (1.05,0.4) {$U^\dg$};
\draw node at (1.05,-0.4) {$U^\dg$};
\draw (1.24, 0.4) -- (1.44,0.4);
\draw (1.24, -0.4) -- (1.44,-0.4);
\end{tikzpicture}
=
\begin{tikzpicture}[baseline=(current  bounding  box.center)]
\draw (0.1, 0.4) -- (0.3,0.4);
\draw (0.1, -0.4) -- (0.3,-0.4);

\draw (0.3, -0.4) -- (0.7,0.4);
\draw (0.3, 0.4) -- (0.45,0.1);
\draw (0.55,-0.1) -- (0.7,-0.4);

\draw (0.7, 0.4) -- (0.9,0.4);
\draw (0.7, -0.4) -- (0.9,-0.4);
\end{tikzpicture}
= W
\end{equation}
Thus when $J_2=0$ and $J=K$, the Hamiltonian has an emergent $SU(3)$ symmetry group.~\footnote{Note that the center of $SU(3)$, which is ${\bf Z}_3$ generated by $e^{i2\pi/3}\text{Id}$, acts trivially on the local degrees of freedom, so technically the symmetry group is $SU(3)/{\bf Z}_3$.  However, this subtlety is not important for our results.}  

For another perspective on the $SU(3)$-symmetric point, we follow reference~\cite{Niesen2017}, noting that, when written in the basis $\{|x\rangle,|y\rangle,|z\rangle\}$, the three components of ${\bf S}$ and five components of ${\bf Q}_v$ are given by the eight Gell-Mann matrices~\cite{GellMann1962}, generators of $SU(3)$.  Thus at the point $J_2=0$, $J=K$, the Hamiltonian is given (up to a constant) by $(J/2)\sum_{\langle ij\rangle}\bm\lambda_{i}\cdot\bm\lambda_{j}$.  This Hamiltonian is an $SU(3)$ analogue of the spin-1/2 Heisenberg model: we simply replace the vector $\bm{\sigma}$ of $SU(2)$ generators by vector $\bm{\lambda}$ of $SU(3)$ generators. 

From this formulation of $H$, it follows that at the $SU(3)$ point, spin-dipole and spin-quadrupole operators must have equal correlation length.  As a result, nematic order cannot extend to $K<J$, while magnetic order cannot extend to $K>J$.  Two possibilities remain: either there is a direct transition between the two orders at precisely the $SU(3)$-symmetric point, or there is an intermediate phase.

\begin{figure*}
\centering
\includegraphics[width=\textwidth]{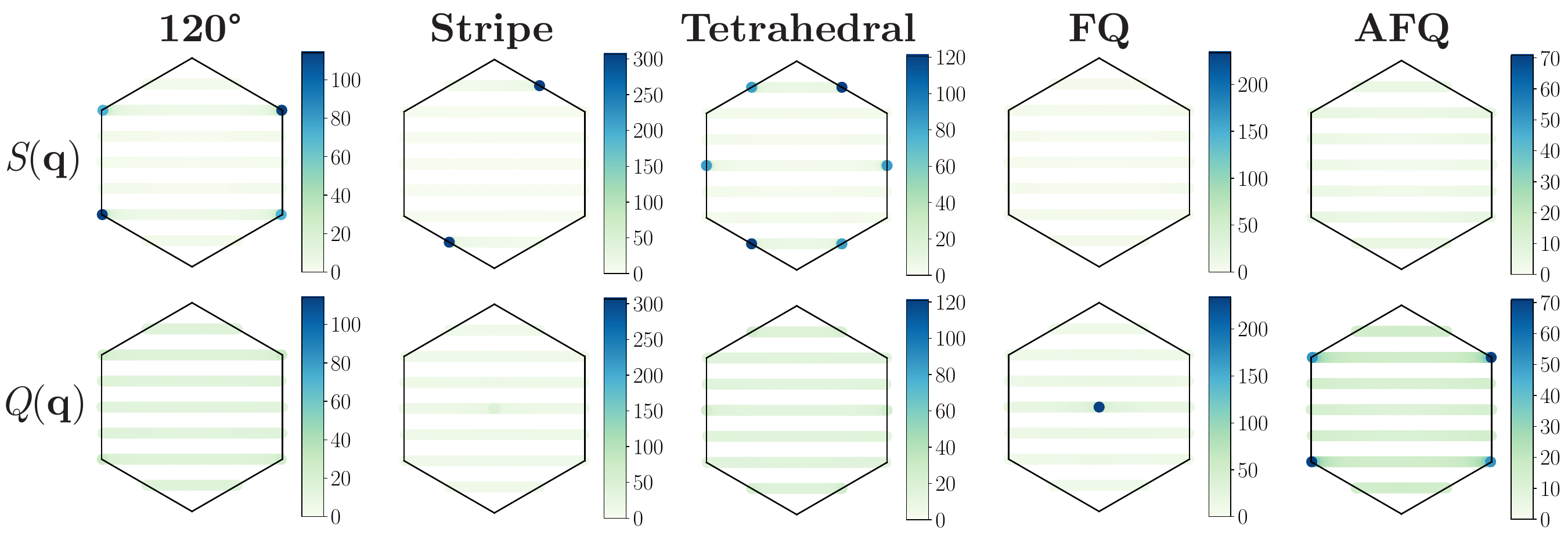}
\caption{(Color online) Top row: spin structure factor $S({\bf q})$ on allowed momentum cuts at one representative point in each phase: respectively, $(J_2/J,K/J)=(0,0.2)$, $(0.3, -0.1)$, $(0.3, 0.2)$, $(0.1, -1.0)$, and $(0.1,1.0)$.  Structure factors are computed using correlations up to 50 unit cells along the cylinder.  Bottom row: corresponding quadrupole structure factors, $Q({\bf q})$.  Note that the scale indicated in the colorbar is consistent between the two structure factors for each phase, clearly indicating whether dipolar or quadrupolar order is dominant, but is different between phases.  All structure factors are computed at bond dimension $\chi=2000$, for which the lowest-energy MPS can spontaneously break spin-rotation symmetry, as discussed in Appendix~\ref{appendix:ent_gap}; this symmetry-breaking leads to the very large peaks in the structure factors.
\label{fig:data_summary_1}}
\end{figure*}

\subsection{Past work}

The \mbox{spin-1} Hamiltonian on the triangular lattice with nearest-neighbor bilinear and biquadratic interactions, but no longer-ranged interactions, has been studied extensively, and there is a broad consensus on the ground state phase diagram.  Including both positive and negative $J$ and $K$, there are four phases: FM, 120$^\circ$ antiferromagnet (AFM), FQ, and AFQ; of these, the FM order exists only for $J<0$, a case we do not consider in the present study.  We now briefly summarize the papers in which these results are presented.

Following the discovery of a non-magnetic ground state in NiGa$_2$S$_4$~\cite{Nakatsuji2005}, Tsunetsugu and Arikawa considered the biquadratic interactions and showed using mean field theory (MFT) that the $K>J>0$ model has an AFQ ground state~\cite{Tsunetsugu2006, Tsunetsugu2007}.  Likewise, Bhattacharjee \emph{et al.} showed that $K<0$, $|K|\gtrsim 1.14 J>0$ gives rise to a FQ ground state in MFT~\cite{Bhattacharjee2006}; they found a first-order transition to 120$^\circ$ magnetic order at $K\approx -1.14 J$.  These works also considered the low-energy excitation spectrum, finding $T^2$ specific heat in both quadrupolar phases, in agreement with the experiments on NiGa$_2$S$_4$.  The low-lying excitations were also addressed using similar mean-field methods by Li \emph{et al.}~\cite{Li2007} and using a continuum field-theory approach by Smerald and Shannon~\cite{Smerald2013}.

A number of works have considered the full $J$-$K$ phase diagram, using a variety of methods. L\"{a}uchli \emph{et al.} showed that both MFT and exact diagonalization (ED) on clusters up to 21 sites give the four phases listed above; however, in the ED calculation the FQ order is stabilized relative to AFM order, so the transition occurs at $K\approx -0.4J$~\cite{Lauchli2006}.  They also used flavor-wave theory~\cite{Papanicolaou1988} to study excitations, and they considered the effects of applied external magnetic fields.  Similar results to the ED were obtained by Moreno-Cardoner \emph{et al.} using cluster MFT with clusters of at least nine sites~\cite{Moreno-Cardoner2014}.  More recently, Niesen and Corboz studied the model using infinite projected entangled pair state (iPEPS) simulations~\cite{Niesen2017,Niesen2018}.  They again observe the same four phases, finding the FQ to AFM phase transition at $K\approx-0.42J$.

Some numerical studies have also targeted particularly important parameter points.  At the $J=0,K<0$ point, with pure biquadratic interactions, Kaul showed using sign-free quantum Monte Carlo (QMC) simulations that the ground state has FQ order~\cite{Kaul2012}.  At the $SU(3)$-symmetric point, $J=K$, Bauer \emph{et al.} found long-ranged quadrupolar order using a combination of flavor-wave theory, DMRG on finite clusters, and iPEPS simulations with 4- and 9-site unit cells~\cite{Bauer2012}.  Zhang \emph{et al.} likewise found indications of long-ranged nematic order at the $SU(3)$ point using DMRG on finite-length cylinders of circumference 6 and 9~\cite{Zhang2021}.  (In contrast, these two studies disagree on the ground state of the $SU(3)$-symmetric model on the square lattice: Ref.~\cite{Bauer2012} finds long-ranged order, whereas Ref.~\cite{Zhang2021} and other recent works by the same authors~\cite{Hu2019,Hu2020} do not.)

A number of studies have added single-ion anisotropy to the model, a term of the form $D\sum \left(S_i^z\right)^2$, employing methods including MFT~\cite{Bhattacharjee2006}, fermionic parton MFT~\cite{Bieri2012, Serbyn2011}, cluster MFT~\cite{Moreno-Cardoner2014}, Schwinger bosons and DMRG~\cite{Wang2017}, and flavor-wave theory~\cite{Bai2021, Seifert2022}.  These works, as well as a quaternion gauge theory calculation on the pure $J$-$K$ model~\cite{Grover2011}, suggest a spin liquid ground state is possible, though it may require the addition of ring exchange~\cite{Bieri2012} or anisotropy in the spin interactions~\cite{Wang2017,Seifert2022}.

Finally, some works have studied the effects of longer-ranged interactions.  The $J$-$K$-$J_3$ model has been studied using a semiclassical approximation~\cite{Stoudenmire2009}, and with fermionic parton MFT~\cite{Liu2010}, suggesting spin-glass-like behavior and a possible spin liquid ground state, respectively. 
The pure $J$-$J_2$ model has been studied using a Green's function approach, giving the following phases, in order of increasing $J_2$: 120$^\circ$ AFM, disordered, stripe magnetic order, incommensurate magnetic order, and three intercollated 120$^\circ$ orders~\cite{Rubin2010}; the first three of these are predicted in the range of $J_2$ we consider in the present work.  
The only work we are aware of that studies the full $J$-$K$-$J_2$ triangular-lattice model does so using a large-$S$ expansion~\cite{Yu2020}, and does not consider $S=1$ in particular.  The predicted large-$S$ phases are the 120$^\circ$ AFM when $J$ dominates, stripe order for sufficiently large $K<0$ and $J_2>J/8-9|K|/16$, and tetrahedral order for $K>0$ and $J_2>J/8-5K/48$.  The authors suggest possible quantum-disordered phases for small $S$ for both signs of $K$, near the boundaries of the AFM phase.


\begin{figure*}
\centering
\includegraphics[width=\textwidth]{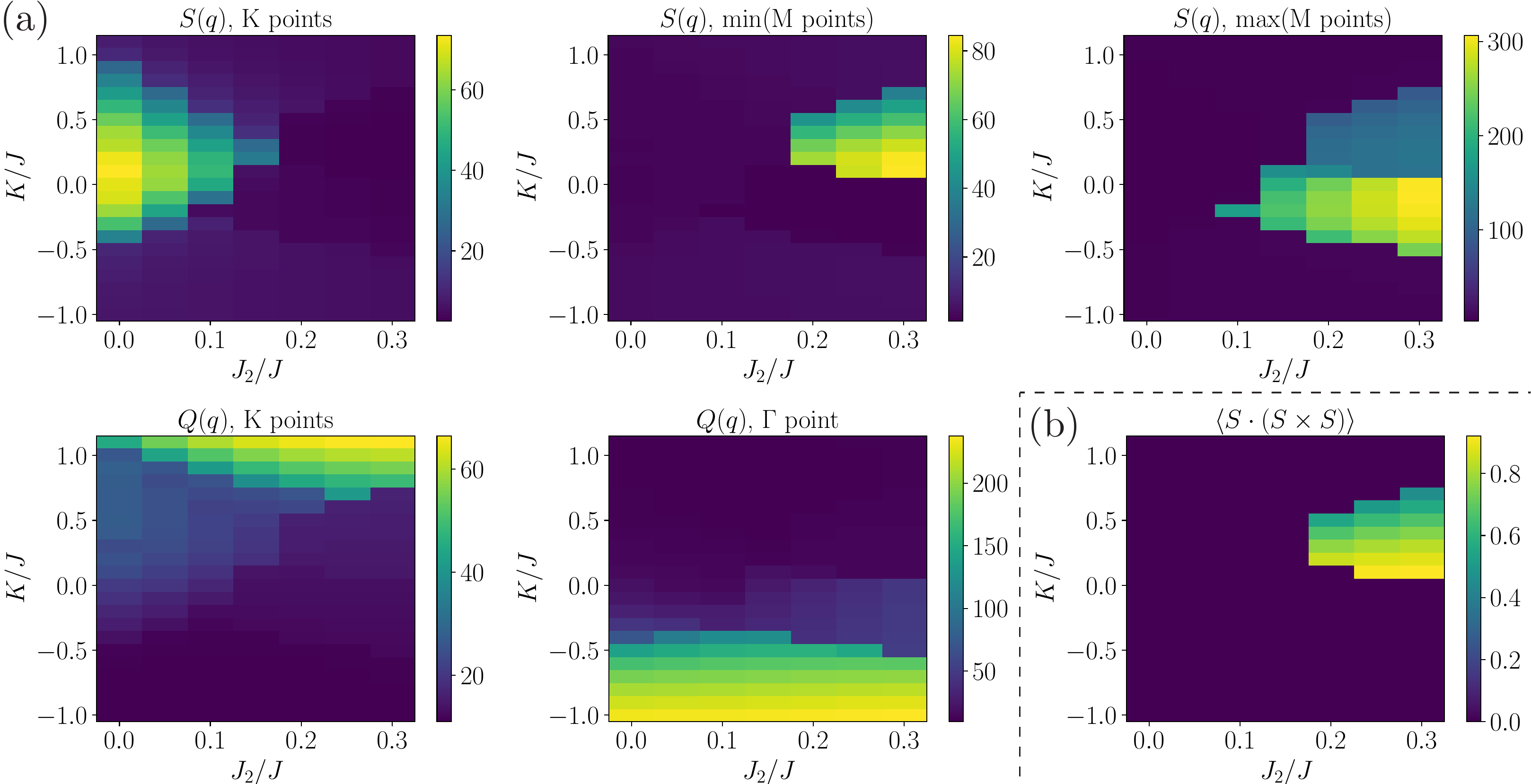}
\caption{(Color online) (a) Height of spin and quadrupole structure factors at specific high-symmetry points in the Brillouin zone (see Fig.~\ref{fig:lattice}), as a function of $J_2/J$ and $K/J$.  The five ordered phases can be visually identified by these data, as explained in the main text.  (b) Scalar chiral order parameter, which is nonzero for tetrahedral magnetic order.
\label{fig:data_summary_2}}
\end{figure*}

\section{Phase diagram and data:\label{sec:phase_diagram}} 

The phase diagram we find is summarized in Figure~\ref{fig:PD}.  In short, we find the expected ordered phases: $120^\circ$ AFM, stripe, and tetrahedral magnetic orders from the classical limit, as well as the expected nematic FQ and AFQ phases.  Our data are plausibly consistent with a disordered phase, but more likely no such phase is present.  Our data suggest the transition between AFM and AFQ orders may be continuous, which is contrary to the conclusions of some past works discussed above.  

Here we present key numerical data on which our summary phase diagram is based and explain how we identify each phase.  Figure~\ref{fig:data_summary_1} shows the spin and quadrupolar structure factors, as defined in equations~\eqref{eq:def_Sq} and \eqref{eq:def_Qq} above, at one representative parameter point in each phase.  The structure factors are computed using correlation functions up to a distance of 50 rings along the cylinder.  Figure~\ref{fig:data_summary_2} provides a different perspective on the structure factors; we plot the values of $S({\bf q})$ and $Q({\bf q})$ at high-symmetry points in the Brillouin zone across the full parameter space, allowing us to map out the locations of the phases and the transitions between them.  Figure~\ref{fig:data_summary_2} also shows the scalar chiral order parameter $\langle {\bf S}\cdot({\bf S}\times{\bf S})\rangle$.  

The data in both figures are from simulations with a single MPS bond dimension, $\chi=2000$.  The main effect of finite bond dimension is that spin-rotation symmetry is spontaneously broken throughout much of the parameter space we study; this symmetry-breaking seems to violate the Mermin-Wagner theorem, which says that a continuous symmetry cannot be spontaneously broken in the ground state of a one-dimensional quantum system such as a finite-circumference cylinder.  However, as we explain in Appendix~\ref{appendix:ent_gap}, breaking symmetry lowers entanglement and thus can also reduce the energy of variational tensor network states; if this energy benefit exceeds the gap between symmetric and symmetry-broken states, the symmetry will indeed be spontaneously broken.  

Apart from the spontaneous symmetry breaking (SSB), the phase diagram at $\chi=2000$ is qualitatively correct.  Compared with $\chi=1000$, for which we show the equivalent of Figures~\ref{fig:data_summary_1} and \ref{fig:data_summary_2} in the Supplemental Material~\cite{SM}, the only changes are small shifts in phase boundaries and a slightly larger region where spin-rotation symmetry is preserved.  Higher-$\chi$ data along $J_2/J=0$ and $J_2/J=0.1$ cuts through parameter space show that there is likewise no qualitative change as we further increase $\chi$ to 4000.  Specifically, we observe consistent behavior in both the structure factors (Figures~\ref{fig:J2_cuts_Sq_Qq_3} and \ref{fig:J2_cuts_Sq_Qq_1}) and correlation lengths (Figure~\ref{fig:J2_cuts_corr_lens}).

We provide further data in the Supplemental Material~\cite{SM}, including real-space spin correlations, spin-resolved correlation lengths, transfer matrix spectra, and entanglement spectra~\cite{Cincio2013} in each phase; while these are not necessary in order to identify the phases, they provide useful additional confirmation of our conclusions.

In the remainder of this section, we proceed phase-by-phase, using the data in the figures to identify each one.  We also separately discuss the nature of the AFM-AFQ transition; some data in the region around the transition are suggestive of a disordered spin liquid phase, and we explain why we believe this interpretation is not correct.

\subsection{Identification of each phase}

$\mathit{120^\circ}\!\!$ \textit{AFM:} The structure factor in Figure~\ref{fig:data_summary_1}(a) has clear peaks at the corners of the Brillouin zone, consistent with three-sublattice magnetic order.  
The extent of the phase is clearly identifiable in Figure~\ref{fig:data_summary_2}(a) from $S(q)$ at the K points.  We note that the phase includes part of the region $0.04\lesssim J_2/J\lesssim 0.25$, where the ground state for $K=0$ was predicted in Ref.~\cite{Rubin2010} to be disordered; while we cannot entirely rule out a disordered ground state in the full two-dimensional model, we find no indication of such behavior.

\emph{Stripe order:} Stripe magnetic order has spins parallel along one lattice vector, and anti-parallel along the other; see the Supplemental Material~\cite{SM} for a real-space picture.  In momentum space, the stripe order has a peak at one of the three $M$ points, as shown in Figure~\ref{fig:data_summary_1}.  In Figure~\ref{fig:data_summary_2}(a), we identify the phase as the region where the minimum over the three $M$ points is small but the maximum is large.  (Note that when we run our simulations independently at each parameter point, the $M$ point at which ordering occurs is chosen spontaneously and is different at different parameter points, but the magnitude of the order parameter is consistent regardless of the choice.)  

\emph{Tetrahedral order:} The tetrahedral order is a four-sublattice non-coplanar order, featuring spin ordering at all three $M$ points and a nonzero scalar chirality, as measured by the order parameter $\langle {\bf S}\cdot({\bf S}\times{\bf S}\rangle$.  We thus identify the extent of this phase from the region with nonzero scalar chirality in Figure~\ref{fig:data_summary_2}(b), which also corresponds to the region with significant ordering at all three $M$ points as indicated by the minimum of the structure factor height over the three points, Figure~\ref{fig:data_summary_2}(a).

\emph{FQ:} The ferroquadrupolar phase has no spin ordering, so the spin structure factor should be small.  However, there is quadrupolar nematic order with the director parallel on all sites, so that the quadrupolar order parameter should have a peak at ${\bf k}={\bf 0}$.  Precisely this behavior is shown in the structure factors at $(J_2/J,K/J)=(0.1,-1)$ in Figure~\ref{fig:data_summary_1}, and we can see the extent of the phase from $Q(q)$ at the $\Gamma$ point in Figure~\ref{fig:data_summary_2}(a).   
As we explain in the Supplemental Material~\cite{SM}, at each parameter point in this phase we actually find two nearly-degenerate ground states.  These correspond to two different ways of spontaneously breaking symmetry due to competition between energy and entanglement (see Appendix~\ref{appendix:ent_gap}), and the true infinite bond dimension ground state is symmetric and unique.

\emph{AFQ:} This phase will again have a small spin structure factor but a large quadrupolar structure factor, now with peaks at the $K$ points.  We show the structure factors at $(J_2/J,K/J)=(0.1, 1.0)$ in Figure~\ref{fig:data_summary_1}, exhibiting precisely this expected behavior.  The extent of the phase can be seen from $Q(q)$ at the K points in Figure~\ref{fig:data_summary_2}(a).   
However, these data do not indicate the full phase; rather, the visually distinct region in the figure corresponds only to the portion of the phase where the symmetry is spontaneously broken.  The full phase extends to smaller $K$, and we identify the full extent of the phase as the region where the quadrupolar structure factor is largest at the $K$ points and the spin-2 correlation length is larger than the spin-1 correlation length (see Fig.~\ref{fig:J2_cuts_corr_lens}).

\subsection{AFM-AFQ phase transition vs spin liquid} 

Finally, we focus on the phase transition between the two three-sublattice orders, AFQ and 120$^\circ$ AFM.  In two dimensions, spin-rotation is spontaneously broken in the true ground state, so one can simply find where the local spin moment goes to 0 to identify the boundary of the AFM phase as in Ref.~\cite{Niesen2018}; the quadrupolar order will be nonzero even in the AFM phase~\cite{Stoudenmire2009}, but a sharp decrease would still make the phase boundary clear.

In contrast, on the cylinder the true $\chi=\infty$ ground state preserves all symmetries on both sides of the transition.  If the transition is first order, there could be significant ordering even infinitesimally away from the transition and thus a very small energy cost to break the symmetry, so that the energy benefit of reduced entanglement can lead to SSB at even sizable bond dimensions (see Appendix~\ref{appendix:ent_gap}).  However, if the transition is continuous or weakly first order, breaking symmetry in favor of the order on one side (say, AFQ) will impose a significant energy cost due to the resulting lack of short-range correlations corresponding to the other order (correspondingly, AFM).  Thus we would expect symmetry to be preserved starting from relatively modest bond dimensions.

\begin{figure}
\centering
\includegraphics[width=0.48\textwidth]{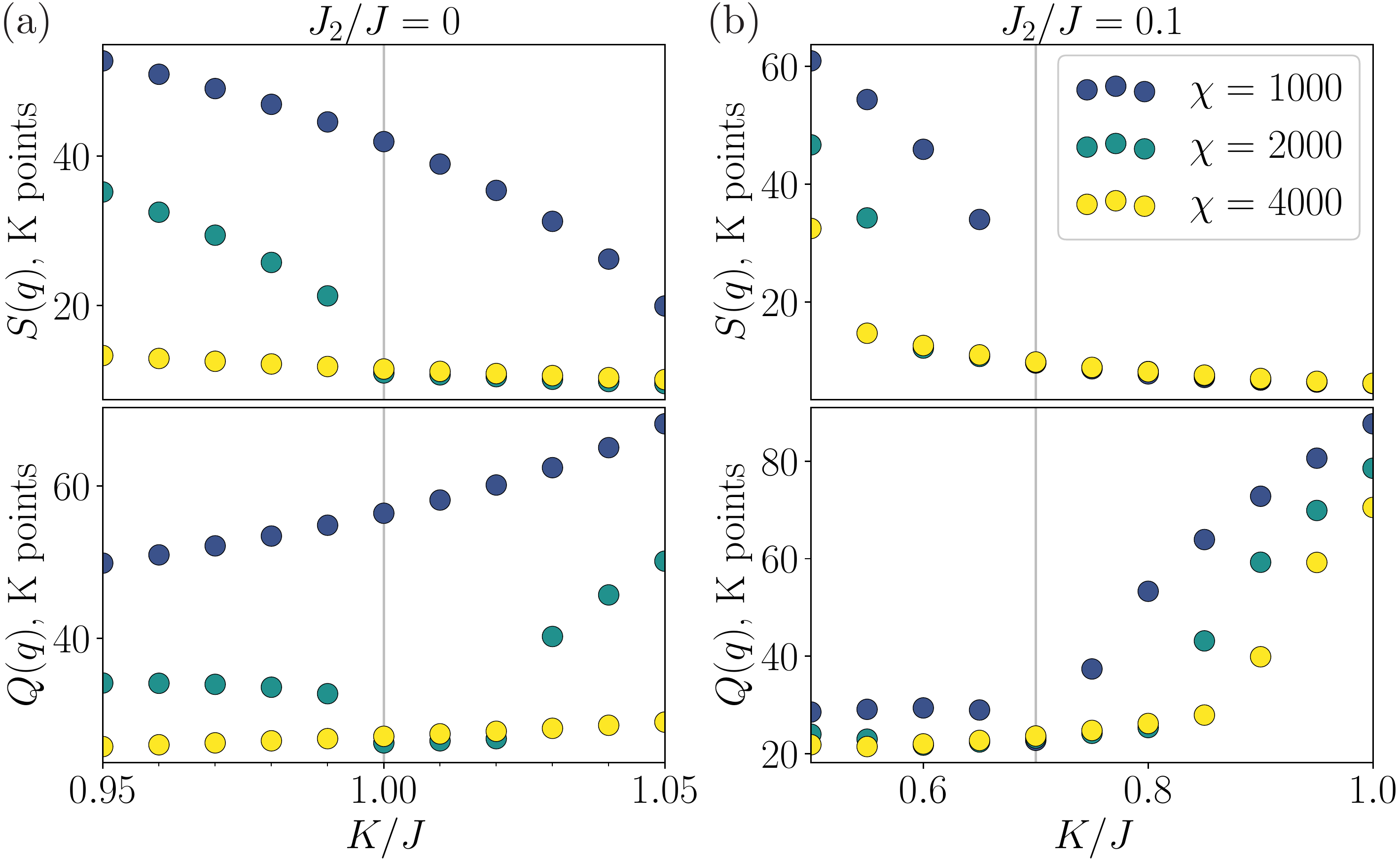}
\caption{(Color online) Height of spin- and quadrupole-structure factors at the $K$ points, as a function of $K/J$ with (a) $J_2=0$ and (b) $J_2/J=0.1$.  Simulations were carried out with a three-ring unit cell, allowing for explicit symmetry-breaking realizing the three-sublattice AFM and AFQ orders.  We observe such symmetry breaking at smaller bond dimension, and the symmetry is restored as $\chi$ increases.
\label{fig:J2_cuts_Sq_Qq_3}}
\end{figure}

\begin{figure}
\centering
\includegraphics[width=0.48\textwidth]{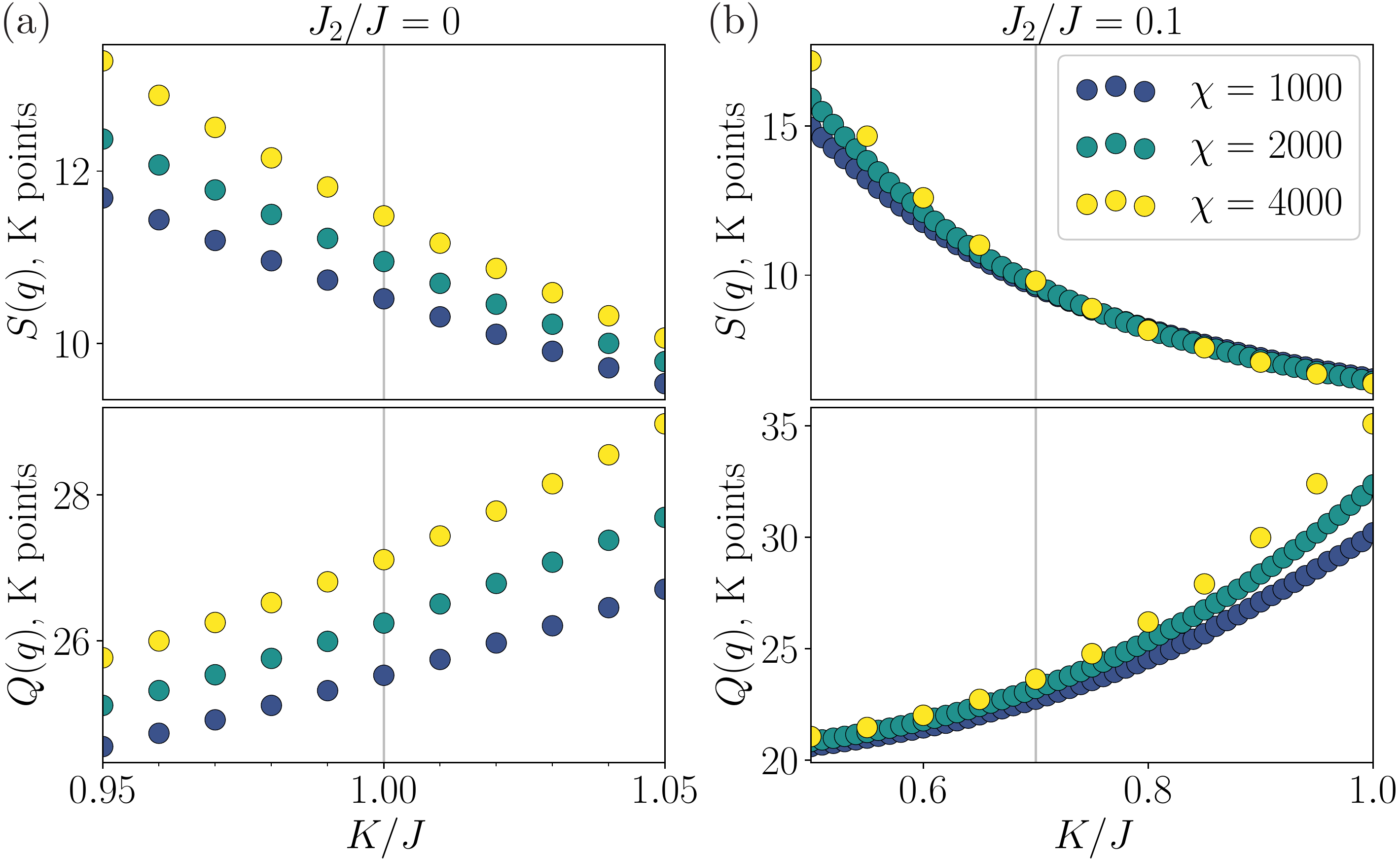}
\caption{(Color online) Height of spin- and quadrupole-structure factors at the $K$ points, as a function of $K/J$ with (a) $J_2=0$ and (b) $J_2/J=0.1$.  Simulations were carried out with a one-ring unit cell, so that explicit symmetry-breaking realizing a three-sublattice order is disallowed. The lowest-energy MPS exactly matches the result for the three-ring unit cell once the bond dimension is large enough in the latter case to restore the symmetry.  In this case the presence of long-range order is not at all clear; however, especially at $J_2/J=0.1$, it appears that each of spin and quadrupole order is converged with bond dimension on one side of $K/J\approx0.7$ and not on the other.
\label{fig:J2_cuts_Sq_Qq_1}}
\end{figure}

As a result, the same numerical data, namely a small region of symmetry-preserving ground states with no long-range order, even at modest bond dimension, is consistent with two possibilities: a disordered region or a direct continuous (or weakly first-order) transition.  On a single finite-circumference cylinder, there is no rigorous way to distinguish these two possibilities.  This is precisely the phenomenology we observe, as shown for the spin-structure factor with a three-ring unit cell in Figure~\ref{fig:J2_cuts_Sq_Qq_3}.  At each sufficiently large bond dimension, there is a region at both $J_2/J=0$ and $J_2/J=0.1$ for which symmetry is preserved and there is no long-range order.  In this region, the ground state exactly matches the results of simulations with a one-ring unit cell, shown in Figure~\ref{fig:J2_cuts_Sq_Qq_1}, for which three-sublattice symmetry-breaking is disallowed.

However, there are several reasons to believe that there is not in fact a disordered phase between the AFM and AFQ orders:

(1) Considering the $J_2/J=0.1$ data, the onset of long-range order shifts significantly when going from bond dimension 1000 to 2000 to 4000.  While this is consistent in principle with an intermediate disordered phase, the apparent phase boundary would likely shift less if there were an underlying phase boundary than if it were simply an artifact of the fact that higher bond dimension decreases the energetic benefit of reducing entanglement by breaking symmetry and hence allows symmetry to be preserved even when further from the phase transition, where the energy cost of breaking symmetry is lower.

(2) We observe similar behavior for the two cuts, $J_2/J=0$ and $J_2/J=0.1$.  As reviewed in Section~\ref{sec:model} above, that there is a direct transition at $J_2=0$ is well-established via a variety of numerical methods, most significantly iPEPS calculations that work directly in the thermodynamic limit~\cite{Niesen2018}.  The symmetry is preserved starting at a lower bond dimension for $J_2/J=0.1$, which does leave open the possibility of a disordered phase, but it could just as well be the case that $J_2$ simply increases the energy cost of breaking the symmetry.

(3) If a disordered phase were indeed a spin liquid or Stiefel liquid, there would be some positive signatures beyond the mere absence of order.  We have checked for topological order by performing flux insertion, which turns out to be $2\pi$-periodic, thus ruling out spin fractionalization.  We also added a scalar chirality term, $J_\chi\sum {\bf S}\cdot({\bf S}\times{\bf S})$, to the Hamiltonian, which would cause a Dirac spin liquid and some Stiefel liquids to transition to a chiral spin liquid (CSL), possibly an $SU(3)_1$ CSL~\cite{Chen2021b}.  Instead, we observe no phase transition with small $J_\chi$, and then a first-order transition to tetrahedral order with larger $J_\chi$; see the Supplemental Material for more~\cite{SM}.  In short, signatures of likely spin liquid phases are not present.

In summary, we cannot rigorously rule out the possibility of an intermediate disordered phase, which could be a spin liquid or a Stiefel liquid.  However, in the absence of truly compelling evidence, we conclude that there is most likely a direct transition between the AFM and AFQ phases.

As part of this argument, we also suggested that the transition is not strongly first order, since if it were we could expect SSB to persist to higher bond dimensions near the transition, especially at $J_2/J=0.1$.  In fact, we can state unequivocally that there is no first-order transition between AFM and AFQ on the YC6 cylinder; we observe numerically that the MPS ground state evolves continuously from one phase to the other, with the overlap between the ground states at adjacent parameter points close to 1. This behavior on the cylinder does not rule out a weakly first-order transition in the two-dimensional model.  

\begin{figure*}
\centering
\includegraphics[width=0.98\textwidth]{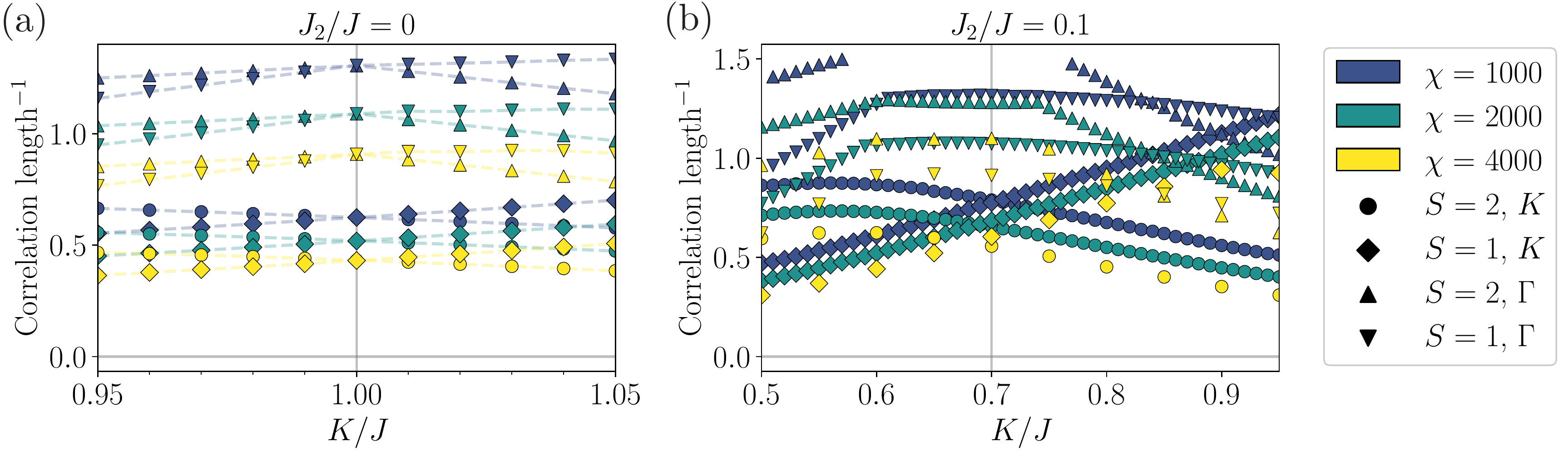}
\caption{(Color online) Inverse correlation length for operators carrying total spin 1 and 2, with momentum corresponding to the $\Gamma$ point (${\bf k}={\bf 0})$ or the $K$ points, for (a) $J_2=0$, and (b) $J_2/J=0.1$.  With $J_2=0$, for each $(S,{\bf k})$, the spin-1 and spin-2 correlation lengths are exactly equal when $K=J$ due to the emergent $SU(3)$ symmetry.  With $J_2/J=0.1$, this is no longer true, and we identify the AFM-AFQ phase transition as the point where the largest spin-1 and largest spin-2 correlation lengths cross. 
\label{fig:J2_cuts_corr_lens}}
\end{figure*}

The continuous transition on the cylinder can be understood in part through correlation lengths for operators with different momentum and spin.  In Figure~\ref{fig:J2_cuts_corr_lens}, we show the correlation lengths, for operators carrying $S=1$ and $S=2$ and corresponding to fluctuations with wave vectors at the $K$ and $\Gamma$ points, across the phase transition both at $J_2/J=0$ and at $J_2/J=0.1$; we briefly explain how to compute spin- and momentum-resolved correlation lengths in the Supplemental Material~\cite{SM} using methods from Refs.~\cite{Pollmann2012,Zauner2015,Gohlke2018}.  The $SU(3)$ symmetry when $J_2=0$, $K=J$ manifests in the exact equality of spin-1 and spin-2 correlation lengths, and this point is the phase transition between the magnetic and nematic orders.  When $J_2/J=0.1$, we likewise identify the parameter point where the longest correlation lengths for spin-1 and spin-2 cross as the phase transition, at $K\approx 0.7J$.

\section{Discussion:\label{sec:discussion}} 

We have numerically studied a highly frustrated model of interacting spin-1 degrees of freedom on the triangular lattice, with antiferromagnetic nearest- and next-nearest-neighbor interactions, and with nearest-neighbor biquadratic interactions.  The biquadratic interactions can be reframed in terms of spin-quadrupole interactions, so that the Hamiltonian leads to competing spin-dipole and spin-quadrupole orders.

We find five ordered phases: 120$^\circ$ antiferromagnetic, stripe, and tetrahedral magnetic orders, and ferroquadrupolar and antiferroquadrupolar nematic orders.  While the high degree of frustration in the model, due to competition both between nearest- and next-nearest-neighbor interaction and between magnetic and nematic order, could plausibly lead to spin liquid or Stiefel liquid ground states, our numerical results suggest that no such disordered state is in fact realized.

A natural follow-up question is how the model could be further modified to produce a spin liquid phase.  The addition of single-ion anisotropy has been suggested as a possible route to spin-liquid behavior, especially with the additional inclusion of ring exchange~\cite{Bieri2012} or anisotropy in the spin interactions~\cite{Wang2017,Seifert2022}.  Another interesting possibility is to explicitly target a nematic spin liquid by studying a model of purely quadrupole interactions, ${\bf Q}_v\cdot{\bf Q}_v$, including contributions beyond nearest-neighbor.

We also suggest that it would be useful to study the model including the next-nearest-neighbor interactions using inherently two-dimensional tensor network simulations, such as with iPEPS.  In particular, since a continuous symmetry can be spontaneously broken in two dimensions even in the true ground state, such simulations would provide further confirmation that there is indeed no disordered state arising between the AFM and AFQ orders.

Finally, our result suggests that the AFM-AFQ transition may be continuous. If this is indeed true on a two-dimensional lattice, the transition is guaranteed to be exotic (non-Landau) for two reasons: (1) The symmetry breaking patterns of the two phases are very different. In particular, the unbroken symmetry of either phase is not the subgroup of the other, so a Landau theory will not describe a direct continuous transition. (2) At the $SU(3)$ symmetric point ($J_2=0$), the system has a Lieb-Schultz-Mattis (LSM) constraint because of the projective representation of $PSU(3)$ per unit cell. The LSM constraint in turn requires any putative field theory of the phase transition to possess certain nontrivial 't Hooft anomaly, a feature clearly absent in any Landau theory. Formulating such a putative theory of continuous AFM-AFQ transition is an interesting problem for future study.


\begin{acknowledgments}
We thank Tim Hsieh and Weicheng Ye for helpful conversations.  DMRG simulations were performed using the TenPy tensor network library~\cite{TenPy2}, which includes significant contributions from Michael Zaletel, Roger Mong, and Frank Pollmann; an updated version of the library is publicly available~\cite{TenPy3}.  Numerical computations were carried out using the Symmetry cluster at Perimeter Institute and the Cedar cluster at Simon Fraser University, through Compute Canada/the Digital Research Alliance of Canada. 
Research at Perimeter Institute is supported in part by the Government of Canada through the Department of Innovation, Science and Economic Development and by the Province of Ontario through the Ministry of Colleges and Universities.  The carbon cost of the simulations reported in this paper was approximately 1.3 metric tons of CO$_2$.
\end{acknowledgments}

\appendix

\section{Spontaneous symmetry breaking and bond dimension\label{appendix:ent_gap}}

Here we briefly address the question of how to interpret the presence or absence of spontaneous symmetry breaking in the MPS ground state at a given bond dimension.  Our goal is to understand the case of a continuous symmetry on an infinite cylinder, but we first illustrate the key idea using a discrete symmetry on a finite one-dimensional system.

Consider, for concreteness, the transverse-field Ising model in one dimension:
\begin{equation}
H = \sum_i J \s_i^z \s_{i+1}^z + h \s_i^x,
\end{equation}
which has a ${\bf Z}_2$ symmetry: $\left(\bigotimes \s^x\right) H \left(\bigotimes \s^x\right)=H$. On a finite chain of length $N$, in the limit $J\gg h$, the two lowest-lying states are nearly degenerate, with splitting proportional to $(h/J)^N$.  Both are eigenstates of the symmetry; in other words, in both the ground state and first excited state, symmetry is not spontaneously broken.  

However, with a finite-bond-dimension MPS, it is possible for the lowest energy variational state to spontaneously break the symmetry.  The bond dimension is the number of Schmidt values allowed to be nonzero on each bipartition of the state, and as the bond dimension increases, the variational space becomes larger and thus the lowest accessible energy becomes lower.  A symmetry-preserving state will have Schmidt values in pairs, and the energy will be determined by the number of pairs kept.  On the other hand, a corresponding symmetry-broken state will achieve the same energy with only half as many Schmidt values, or in other words will have lower energy when the Schmidt rank is the same.  Consequently, at finite bond dimension there is an energy cost to preserving the symmetry, which we will call the ``symmetry-entanglement gap.''  

Symmetry will be spontaneously broken if this gap is larger than the exponential splitting $(h/J)^N$.  Infinite DMRG will thus spontaneously break discrete symmetry in 1D for any bond dimension, since the true splitting is exactly zero and the symmetry-entanglement gap is merely tiny.  Finite DMRG will break the symmetry up to some cutoff bond dimension, beyond which it will be restored.

We now turn to the more complicated case of a 2D model with a continuous symmetry, which is then restricted to a finite-circumference cylinder.  The Mermin-Wagner theorem says that on this quasi-1D system, the true ground state cannot spontaneously break the symmetry.  However, in DMRG simulations such SSB is possible, as we can understand by analogy with the 1D Ising model.  The cylinder circumference plays the role of the 1D chain's length: when it becomes infinite, exact ground state degeneracy allows SSB to occur even in a true ground state, but when it is finite, the true ground state must be symmetry-preserving due to an energy splitting exponentially small in cylinder circumference.  This splitting can be compared with the symmetry-entanglement gap due to finite bond dimension, with the result that even on a finite-circumference cylinder where the true ground state must respect the symmetry, the lowest-energy MPS will break the symmetry for sufficiently small bond dimension.  Furthermore, the requisite bond dimension to preserve the symmetry will increase with cylinder circumference.

Suppose, on the other hand, that in the true two-dimensional ground state the continuous symmetry cannot be spontaneously broken, such as for a topological spin liquid.  This implies an energy gap, $\Delta$, above the symmetric ground state.  If we view the main effect of finite cylinder circumference as restricting to certain momentum cuts in the Brillouin zone (which is a reasonable perspective when the circumference is large, less so when small), the corresponding gap on the cylinder will be comparable or larger.  Hence for the lowest-energy MPS to spontaneously break the symmetry, we require the symmetry-entanglement gap to be larger than $\Delta$.  For any reasonable circumference, $\Delta$ will be substantially larger than the finite-circumference gap of a state with SSB, so that a continuous symmetry will only be broken when the bond dimension is quite small.

To summarize, with sufficiently small bond dimension the lowest-energy MPS on a finite-circumference, infinite-length cylinder will be expected to spontaneously break symmetry, regardless of what happens in the 2D model or the true ground state.  However, if there is long-ranged order in 2D, the threshold bond dimension required to restore the symmetry on the cylinder will be larger than if the 2D ground state is disordered.  Thus as a practical conclusion, we can guess that if a very large bond dimension is required in order to restore the symmetry, long-range order is likey present in 2D.  Conversely, if the simulation setup allows the symmetry to be spontaneously broken but the symmetry is nonetheless respected even with very small bond dimension, the 2D ground state likely has no long-ranged order.


%

\onecolumngrid

\clearpage



\section*{Supplementary material}

## I. REAL-SPACE CORRELATIONS

In Fig. 1, we show the real-space spin and quadrupole correlations in each phase, from which we compute the structure factors shown in Fig. 4 of the main text. The three-sublattice patterns of the spin correlations in the 120° antiferromagnetic (AFM) phase and the quadrupole correlations in the antiferroquadrupolar (AFQ) phase are readily apparent, as is the four-sublattice pattern in the tetrahedral phase. Likewise, we can immediately see the stripe magnetic ordering and the translation-invariant, i.e. $\mathbf{k} = \mathbf{0}$, nature of quadrupole correlations in the ferroquadrupolar (FQ) phase.

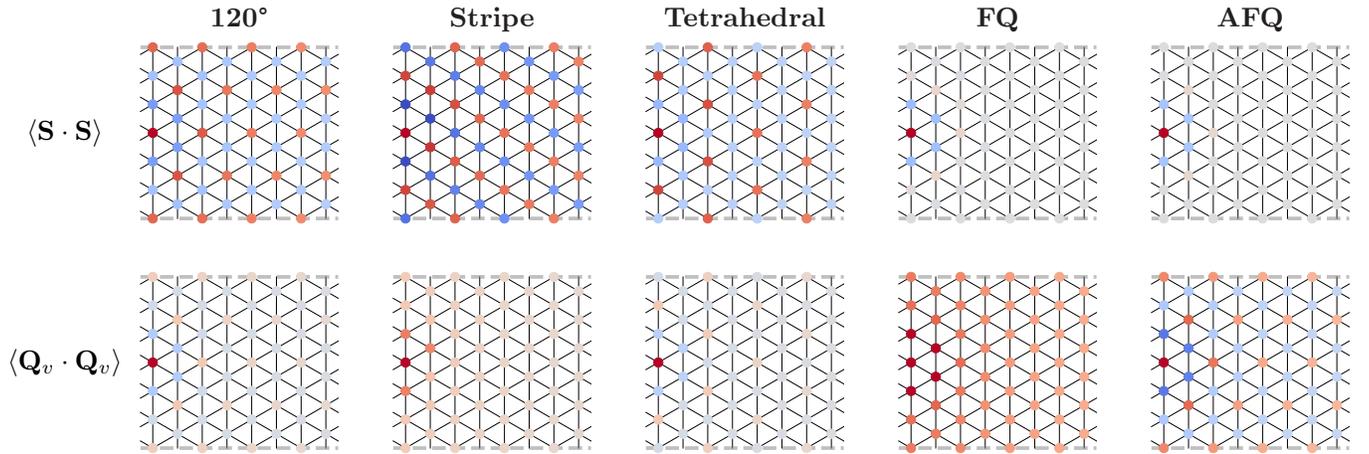

FIG. 1. Real-space spin and quadrupole correlations. For each site shown, the correlation is computed to the site in the center on the left (dark red); red indicates positive correlation, and blue indicates negative correlation.

## II. SMALLER BOND DIMENSION DATA

In Figs. 2 and 3, we show the equivalent to Figs. 4 and 5 of the main text, but for a smaller bond dimension, $\chi = 1000$ rather than $\chi = 2000$. The structure factors in Fig. 2 are qualitatively the same, showing that the nature of the observed phases does not change between these bond dimensions.

The full parameter-space view in Fig. 3 shows two key differences compared with $\chi = 2000$ data shown in the main text. (1) the symmetry-preserving region between 120° AFM and AFQ grows with increasing bond dimension, as also shown in Fig. 6 of the main text. As explained in Appendix A of the main text, the difference arises because when the bond dimension is larger, there is a smaller energy benefit from the reduced entanglement of a symmetry-broken state.

---

* aszasz@perimeterinstitute.ca

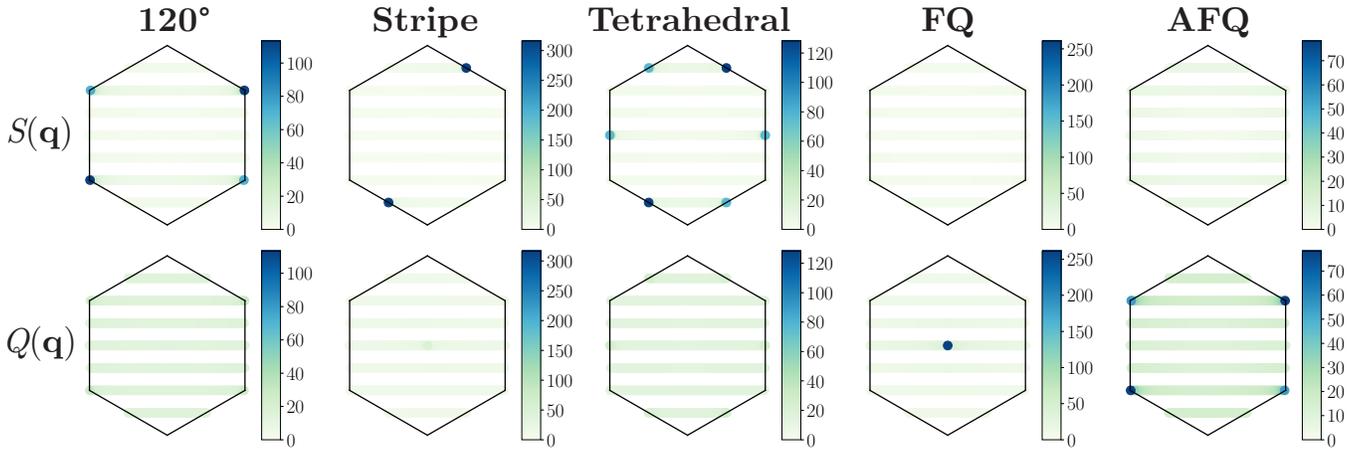

FIG. 2. Equivalent of Fig. 4 of the main text, for bond dimension $\chi = 1000$. The structure factor for each phase is computed at the same parameter point as in that figure.

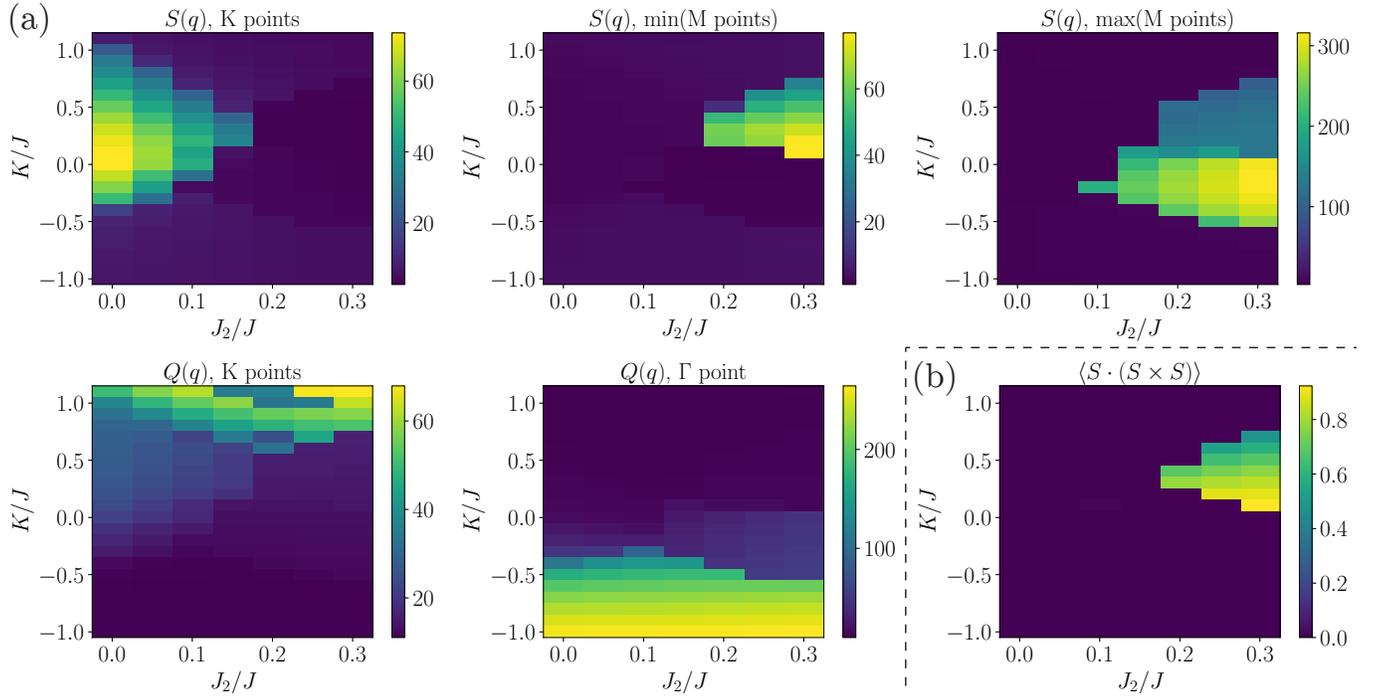

FIG. 3. Equivalent of Fig. 5 of the main text, for bond dimension $\chi = 1000$.

## III. TRANSFER MATRIX SPECTRA AND CORRELATION LENGTHS

In this section, we briefly review the matrix product state (MPS) transfer matrix, and how to label the spectrum of the transfer matrix by quantum numbers/conserved quantitites such as spin and momentum (Sec. III A). We then show data for computed transfer matrix spectra and correlation lengths in each phase (Sec. III B) and with flux insertion (Sec. III D).

### A.  How to compute spin- and momentum-resolved transfer matrix spectra & correlation lengths

Consider an infinite MPS with a single-site unit cell, thus specified by the tensor $A^\sigma_{\alpha\beta}$ where $\sigma$ labels the physical degree of freedom and $\alpha$ and $\beta$ the auxiliary legs. We then construct the transfer matrix, $T$, as follows:

$$\alpha - A - \beta \quad \to T = \begin{matrix} -A^* - \\ | \\ -A- \end{matrix} \qquad (1)$$

We can compute the eigenvalue spectrum of $T$, viewed as a matrix from the right two legs to the left two legs; denote this spectrum by $\{\lambda\}$. A correlation function between two sites with separation $r$ will involve multiplication by $r-1$ copies of the transfer matrix, hence depending on the operator will look like $\sum_i c_i \lambda_i^r$ for some coefficients $c_i$. Thus the decay for large $r$ will go as $\lambda^r = e^{-r/\xi}$ where the correlation length $\xi$ is $\mathrm{Re}[-1/\log(\lambda)]$ for $\lambda$ the largest eigenvalue such that $c_i \neq 0$ for the particular operator involved in the correlation function; $\xi$ is measured in unit cells. In other words, $\{-1/\log(\lambda)\}$ gives the full set of possible correlation lengths for different operators. As explained in the SM of Ref. [1], Sec. IV A, the $\lambda$ can be labeled by conserved quantities such as spin, allowing us to find a spin-resolved correlation length.

It is generally true that the correlation length for an operator is inversely proportional to the energy gap to the excitation created by that operator. We thus can also use the transfer matrix eigenvalues to learn about the excitation spectrum, which is approximately proportional to $\{\mathrm{Re}[-\log(\lambda)]\}$. This latter set is what we refer to as the "transfer matrix spectrum" in the remainder of the Supplemental Material and plot in several figures below.

As argued in Ref. [2], we can extract even more information from the transfer matrix spectrum. We have thus far used only the real part of $\log(\lambda)$ to get the correlation length and excitation spectrum; we now consider the imaginary part. Specifically, let $\lambda = \exp(-1/\xi + ik)$, so that a corresponding correlation function will go as $\exp(-r/\xi)\exp(ikr)$; evidently, $k = \mathrm{Im}[\log(\lambda)]$ plays the role of momentum corresponding to translation between MPS unit cells.

We now move to the case of the triangular lattice on an infinite cylinder. The momentum extracted from the imaginary part of $\log(\lambda)$ corresponds again to translation between unit cells. This translation is along **a** in Fig. 4, so the derived momentum is along $\mathbf{k}_a$. However, we can also get the component of momentum along $\mathbf{k}_b$. To do so, we use the gauge-invariance of the MPS. Following Ref. [3], we suppose first that we have a translation-invariant MPS with a single-site unit cell, and that it describes a wave function with a global on-site symmetry: $\mathcal{O}^{\otimes N}|\psi\rangle = e^{i\theta}|\psi\rangle$ where $\theta$ is a physically irrelevant overall sign; since $\theta$ could exist in principle, we must keep it. This symmetry operation can be implemented by applying $\mathcal{O}$ on the physical leg of the MPS tensor, and to get the same $|\psi\rangle$ at the end, the most general possibility is:

$$\begin{matrix} \mathcal{O} \\ | \\ -A- \end{matrix} = e^{i\phi} -U-A-U^\dagger- \qquad (2)$$

for some unitary $U$, which will cancel between adjacent unit cells. ($\phi$ is related to the overall phase $\theta$.) Note the following convention: all matrices (rank-2 tensors) are drawn such that, if the legs are horizontal, the one on the left corresponds to the row index and the one on the right to the column index, and if vertical, the downwards leg corresponds to rows and upwards to columns.

We assume the MPS is normalized, so the largest eigenvalue of the transfer matrix is 1. If it is specifically right-normalized, we can extract $U$ numerically:

$$\begin{matrix} -A^*- \\ | \\ -A- \end{matrix}\!\!\!) = \;) \;\Rightarrow\; \begin{matrix} -A^*- \\ \mathcal{O} \\ -A-U \end{matrix}\!\!\!) = e^{i\phi} \begin{matrix} -A^*- \\ | \\ -U-A-U^\dagger-U \end{matrix}\!\!\!) = e^{i\phi}\; -U\;) \qquad (3)$$

The object in the dashed box is called the "mixed transfer matrix." Evidently $U$ is an eigenvector of the mixed transfer matrix, with eigenvalue $e^{i\phi}$. We can do something analogous if we replace the dotted box, which is the leading eigenvector of the standard transfer matrix $T$, by a different eigenvector of $T$ with eigenvalue $\lambda$:

$$\cdots = \lambda \cdots \Rightarrow \cdots = \lambda e^{i\phi} \cdots = \lambda e^{i\phi} \cdots \qquad (4)$$

Evidently, there is a precise one-to-one correspondence between eigenpairs of the original and the mixed transfer matrices; the latter are given by multiplying the eigenvalues of the former by $e^{i\phi}$ and the eigenvectors by $U$.

Since $\mathcal{O}$ is a symmetry, there is a corresponding conserved quantity, and we can assign a charge (quantum number) to each eigenpair of the mixed transfer matrix, hence to each eigenvalue of the original transfer matrix since they are in bijection.

In the case of the cylinder, we can block all the sites in a ring into a single super-site. Then translation by one site around the cylinder is an on-site global symmetry for these super-sites, and we have (shown for a circumference of three sites as an example):

$$\cdots = e^{i\phi} \cdots \qquad (5)$$

The quantum numbers associated to the transfer matrix eigenpairs from the corresponding mixed transfer matrix will be $e^{ik_b}$ where $k_b$ is the component of momentum along $\mathbf{k}_b$ in Fig. 4.

As argued in Ref. [4], if $U$ is diagonal, each entry will have a phase $e^{iq}$ for $q = (2\pi/L_y)n$, and this indicates the momentum $q$ associated to each index value of the virtual leg. Then if the transfer matrix eigenvector $\Lambda$ in Eq. (4) indeed has a well-defined momentum $k_b$, it will only have nonzero entries in matrix elements where the difference between the momentum of the row index, $q_r$, and that of the column index, $q_c$, is $k_b = q_r - q_c$. We then have $U\Lambda U^\dagger = e^{ik_b}\Lambda$, and can therefore extract $e^{ik_b}$ by comparison with $\Lambda$.

But what if $U$ is not diagonal? Let us assume the tensor $A$ is such that the $U$ extracted by as the leading eigenvector of the mixed transfer matrix is indeed diagonal. We then transform our MPS by conjugating $A$ by some unitary $\tilde{U}$. This has no effect on the actual physical state corresponding to the MPS, but it will change the leading eigenvector of the mixed transfer matrix to $U' \equiv \tilde{U}U\tilde{U}^\dagger$, which in general is no longer diagonal:

$$\cdots = e^{i\phi} \cdots \Rightarrow \cdots = e^{i\phi} \cdots \qquad (6)$$

Here the diamond indicates that $U$ is diagonal. Likewise, the eigenvector $\Lambda$ transforms as

$$\cdots = \lambda \cdots \Rightarrow \cdots = \lambda \cdots \qquad (7)$$

so $\Lambda \mapsto \Lambda' \equiv \tilde{U}\Lambda\tilde{U}^\dagger$. Note that, on the other hand, the eigenvalue $\lambda$ is unchanged—this is essential, since it means the transfer matrix spectrum is gauge-invariant! Also note that when $\Lambda = \mathrm{Id}$, conjugation by $\tilde{U}$ has no effect, showing that this $\tilde{U}$ gauge is not fixed by right-normalization.

In that case, $U'\Lambda'U'^\dagger = \tilde{U}U\Lambda U^\dagger \tilde{U}^\dagger = e^{ik_b}\tilde{U}\Lambda\tilde{U}^\dagger = e^{ik_b}\Lambda'$. We conclude that we can extract $k_b$ even when the MPS is in a gauge where $U$ is not diagonal, by comparing $U'\Lambda'U'^\dagger$ with $\Lambda'$; they will differ precisely by an overall factor of $e^{ik_b}$. In practice, the easiest way to extract this phase factor is to normalize the eigenvector $\Lambda'$, then compute the dot product between $\Lambda'$ and $U'\Lambda'U'^\dagger$. In other words, for $U$ and $\Lambda$ defined by

$$\cdots = e^{i\phi} \cdots , \qquad \cdots = \lambda \cdots , \qquad (8)$$

regardless of the gauge of the MPS, we can compute the quantum number $k_b$ corresponding to the eigenpair $(\lambda, \Lambda)$ as

$$e^{ik_b} = \quad\quad\quad = \Lambda \cdot U \Lambda U^\dagger \quad\quad\quad (9)$$

Note that alternatively, upon finding $U$, one could explicitly diagonalize it to obtain $\tilde{U}$ and use the original formulation from Ref. [4].

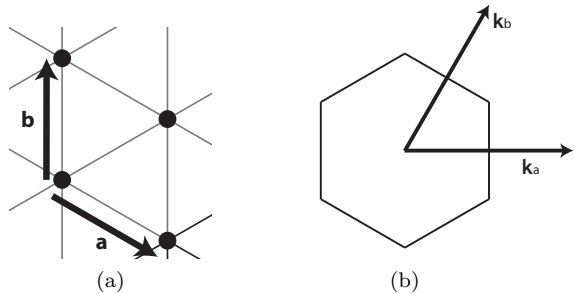

FIG. 4. (a) Lattice vectors **a** and **b** on the triangular lattice. (b) Corresponding reciprocal lattice vectors and the first Brillouin zone.

Finally, we can now associate to each transfer matrix eigenvalue a precise momentum. For each eigenvalue $\lambda$, we have seen how to get the components of **k** along both $\mathbf{k}_a$, from the imaginary part of $\log \lambda$, and $\mathbf{k}_b$, from the mixed transfer matrix. Since these momentum components are independent, we then get the total momentum $\mathbf{k} = n_a \mathbf{k}_a + n_b \mathbf{k}_b$ where $n_a$ and $n_b$ are each in $[-0.5, 0.5)$. (Note that these are found from the scalar $k_a$ and $k_b$ by $n_\alpha = k_\alpha/2\pi$.)

In the figures below, we show the transfer matrix spectrum plotted both vs $n_a$ and vs $n_b$, thus allowing each point to be precisely located in the Brillouin zone. Some key points include the $K$ points, at $(n_a, n_b) = \pm(1/3, 1/3)$, the $\Gamma$ point at $(0,0)$, and the $M$ points at $(0, 1/2)$, $(1/2, 0)$, and $(1/2, -1/2)$. (Note that in finding $n_b$, the sign will depend on the direction of the swap in Eq. (5), and if the wrong choice is accidentally made, it can be corrected by noting that in certain phases the largest eigenvalues should be at the $K$ points.)

### B. TM spectrum and correlation lengths in each phase

We use the method explained in Section III A to compute the transfer matrix spectrum, resolved by spin and momentum, in each of the five ordered phases. We show the spectrum at a respresentative point in each phase in Fig. 5. The parameter points are the same as in Fig. 4 of the main text, except for in the tetrahedral phase where we use $(0.3, 0.5)$. Recalling that $\mathbf{k} = n_a \mathbf{k}_a + n_b \mathbf{k}_b$, for $\mathbf{k}_a$ and $\mathbf{k}_b$ as shown in Fig. 4, the left panel for each parameter point shows the spectrum vs $n_a$ and the right vs $n_b$. The color indicates the quantum number $s_z$, and we can identify spin multiplets by looking for degeneracy of points with quantum numbers $-s$ to $s$ all at the same $n_a$ and $n_b$. The multiplets are not identified in Fig. 5, but are used to construct Fig. 8 of the main text.

While the full transfer matrix spectra provide a detailed look at low-energy excitations, we can gain additional insight by ignoring the momentum, and finding the longest correlation length for each spin quantum number $S_z$. We show $\xi$ vs $S_z$ for a representative point in each phase in Fig. 6. If spin-rotation symmetry is preserved, we expect to see that the longest correlation length corresponds to a spin multiplet, i.e. that the correlation lengths are equal for $S_z$ from $-s$ to $s$. This is true in the 120° and AFQ phases with a 1-ring unit cell, since that unit cell prevents SSB corresponding to a three-sublattice order. In those phases, but with a 3-ring unit cell, and in all other phases (for which the spin- or quadrupole-order is allowed with a 1-ring unit cell), we can directly see the breaking of spin-rotation symmetry. We expect that the $S_z$ value with the largest correlation length is the total spin of the multiplet that would be found if the symmetry were restored. We also find two nearly degenerate states for the FQ phase, with the largest correlation lengths having different $S_z$; see Section IV below for an explanation.

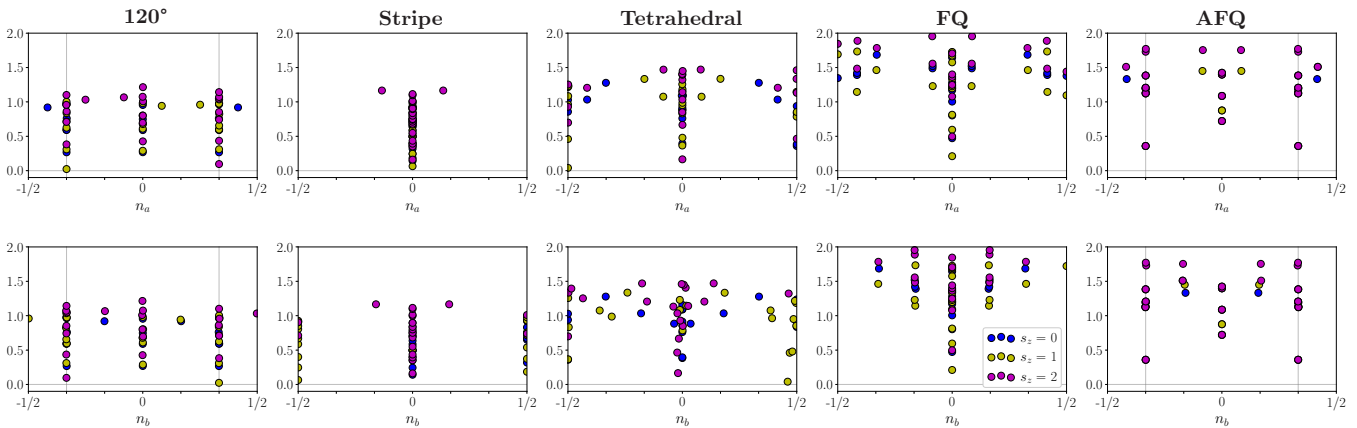

FIG. 5. Spin- and momentum-resolved transfer matrix spectra for a representative point in each phase, computed with bond dimension $\chi = 2000$. The two figures for each parameter point give the momentum of each transfer matrix eigenvalue as $\mathbf{k} = n_a \mathbf{k}_a + n_b \mathbf{k}_b$. Notes: (1) The $K$ points are at $(n_a, n_b) = \pm(1/3, 1/3)$, and the $M$ poins are at $(0, \pm 1/2)$, $(\pm 1/2, 0)$, and $\pm(1/2, -1/2)$; vertical lines are a guide to the eye. (2) All data in the figures come from simulations with a one-ring unit cell. (3) The $n_b$ plot for the tetrahedral order does not show clean quantization because (at this bond dimension) the wavefunction breaks translation symmetry around the cylinder. (4) The gap at the top of some figures is because we compute a fixed number of transfer matrix eigenvalues for each $s_z$, and they are denser at the bottom of the spectrum in these phases.

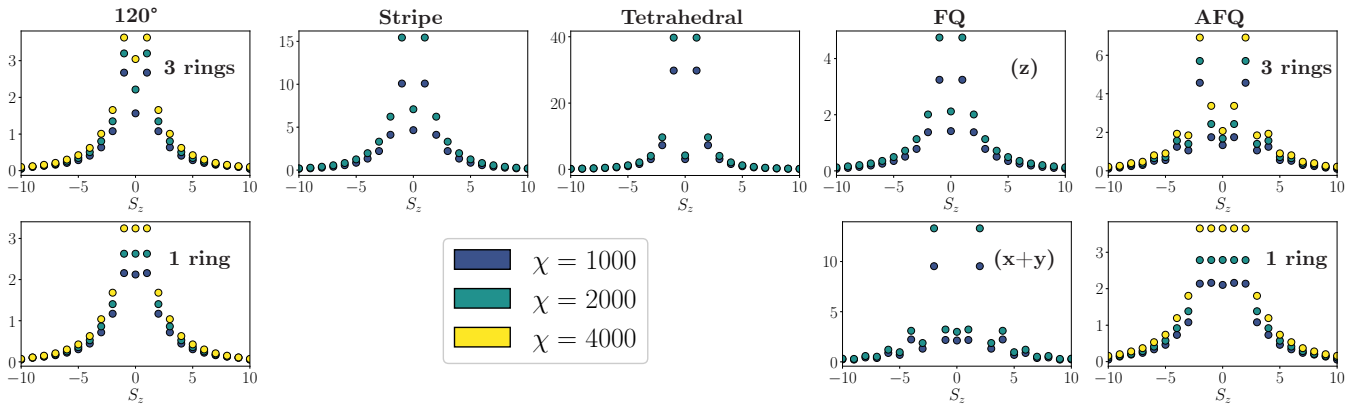

FIG. 6. Correlation length (measured in sites along the length of the cylinder) for operators carrying different spin quantum numbers, $S_z$, at a representative point in each phase. We show the result for both one- and three-ring unit cells in the AFM and AFQ phases; spin-rotation is manifestly broken with the three-ring unit cell, and preserved with the one-ring unit cell. We also show the result for two nearly-degenerate states in the FQ phase; the labels $(x+y)$ and $(z)$ are explained in Section IV below.

### C. TM spectrum and correlation lengths at the AFM-AFQ transition

Similarly, we look briefly at the transfer matrix spectra and correlation lengths right at the transition between AFM and AFQ orders. When $J_2 = 0$, this occurs precisely at $K = 1$, while at $J_2 = 0.1J$, the transition is at $K \approx 0.7J$. We show the transfer matrix spectrum and spin-resolved correlation length at these two parameter points in Figs. 7 and 8, respectively. As indicated in Fig. 8 of the main text, in both cases the largest correlation length is actually 16-fold degenerate, corresponding to one spin-2 multiplet and one spin-1 multiplet at each of the two $K$ points. At $J_2 = 0$, $J = K$, there is likewise a degeneracy between spin-2 and spin-1 multiplets through the whole spectrum due to the $SU(3)$ symmetry.

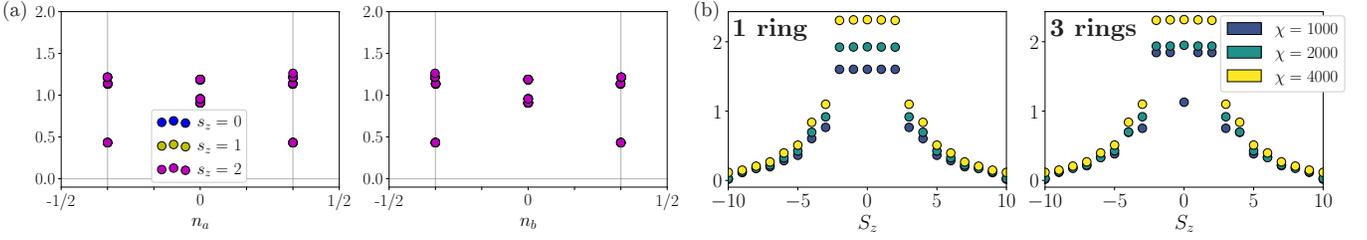

FIG. 7. (a) Transfer matrix spectrum vs momentum, at $J_2 = 0$, $J = K$, computed with $\chi = 4000$. This parameter point corresponds to the phase transition between AFM and AFQ orders, and has an emergent $SU(3)$ symmetry. The spectrum has an exact degeneracy between spin-1 and spin-2 multiplets, which is why so few points are visible compared with the spectra in Fig. 5. (b) Longest correlation length for each spin quantum number, $S_z$, computed both with one- and three-ring unit cells. A spin-2 multiplet is clearly visible from the 5-fold degeneracy, but in fact the largest correlation-length is 16-fold degenerate, for two spin-2 multiplets and two spin-1 multiplets. For the three-ring unit cell, the symmetry-breaking at low bond dimension is apparent.

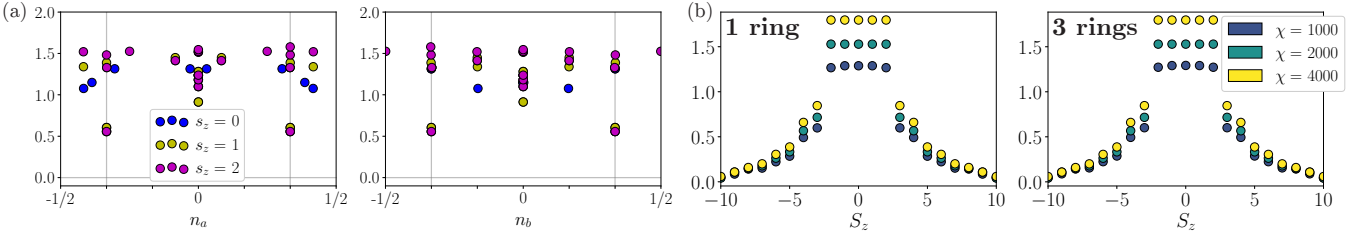

FIG. 8. (a) Transfer matrix spectrum vs momentum, at $J_2/J = 0.1$, $K/J = 0.7$, near the AFM-AFQ transition, with $\chi = 4000$. There is an approximate degeneracy between the lowest spin-1 multiplet and the lowest spin-2 multiplet at the $K$ point; it would be exact on the transition line. Unlike at the $SU(3)$ point, this degeneracy does not extend through the full spectrum. (b) Correlation length vs $S_z$, showing similar behavior to that at the $SU(3)$ point, although here the symmetry is not broken for the three-ring unit cell, even at $\chi = 1000$.

### D. Transfer matrix spectra with flux insertion

As mentioned above, the transfer matrix spectrum acts as a proxy for the excitation spectrum. However, the information about the excitation spectrum that we can gain is limited by the quantization of $k_y$ due to the finite cylinder circumference. To overcome this difficulty and get a much more complete picture of the excitation spectrum, we make use of flux insertion through the cylinder.

When we insert spin-flux $\theta$, the result is to modify the spin interactions across the periodic boundary of the cylinder. $S_{L-1}^+ S_0^-$ becomes $e^{-i\theta} S_{L-1}^+ S_0^-$ and $S_{L-1}^- S_0^+$ becomes $e^{i\theta} S_{L-1}^- S_0^+$, with $S^z S^z$ unchanged; we still have the original periodic boundary condition that $\mathbf{S}_L = \mathbf{S}_0$. The flux seems to break translation symmetry around the cylinder, since the terms across the boundary have extra phases compared with the rest. We can restore the symmetry by a change of variables: $S_j^+ \mapsto e^{i\theta j/L} S_j^+$, $S_j^- \mapsto e^{-i\theta j/L} S_j^-$, so that every $\mathbf{S_j} \cdot \mathbf{S_{j'}}$ interaction becomes $S_\mathbf{j}^z S_\mathbf{j'}^z + (e^{i\theta \Delta_y/L} S_\mathbf{j}^- S_\mathbf{j'}^+ + e^{-i\theta \Delta_y/L} S_\mathbf{j}^+ S_\mathbf{j'}^-)/2$, with $\Delta_y = (\mathbf{j'} - \mathbf{j}) \cdot \mathbf{k}_b/(2\pi)$. However, we now have $S_L^+ = e^{i\theta} S_0^+$ and $S_L^- = e^{-i\theta} S_0^-$, in other words twisted boundary conditions. The twisted boundary conditions in turn result in quantized momenta shifted by $\theta/L$, so in tuning $\theta$ from 0 to $2\pi$, the allowed momentum cuts sweep out the full Brillouin zone.

We have performed this flux insertion at several points in the phase diagram. The flux insertion is adiabatic, in the sense that we take a high resolution in $\theta$, and we initialize the simulation for each $\theta$ with the result from the preceding one. We show the results here, and explain the conclusions we draw as a result. Note that the flux insertion explicitly breaks spin-rotation, so splitting of spin-multiplets in the spectrum is expected.

In Fig. 9, we show the spin- and momentum-resolved transfer matrix spectrum at $(J_2/J, K/J) = (0.1, -0.5)$ in the FQ phase. By plotting the results together for the full range of $\theta$, we observe that the complete excitation spectrum has a minimum at the $\Gamma$ point.

In Fig. 10, we show the transfer matrix spectrum computed at $(J_2/J, K/J) = (0.1, 0.7)$, approximately at the AFM-AFQ transition, with a three-ring unit cell. In Fig. 11, we show the result at the same point but using a one-ring unit cell. In both cases, we observe a first-order transition from the zero-flux state to some new state, then back to the original state as flux approaches $2\pi$. In the one-ring case, it appears that correlation length could be diverging as

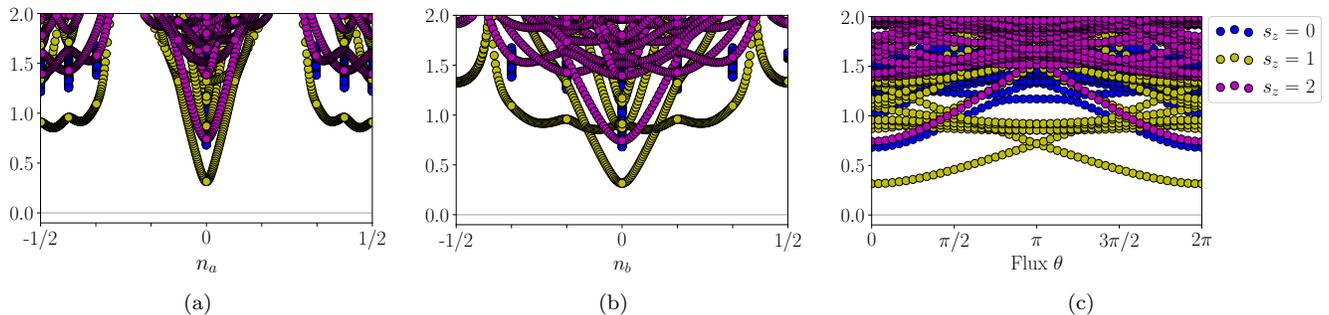

FIG. 9. Transfer matrix spectrum for $(J_2/J, K/J) = (0.1, -0.5)$ in the FQ phase ($z$-type state, see Section IV), using flux insertion through the cylinder to sweep the allowed momentum cuts through the full Brillouin zone, with $\chi = 2000$. We show the spectrum vs (a) $n_a$, (b) $n_b$, and (c) the inserted flux, $\theta$. The lowest energy excitations are evidently at the $\Gamma$ point, $\mathbf{k} = \mathbf{0}$.

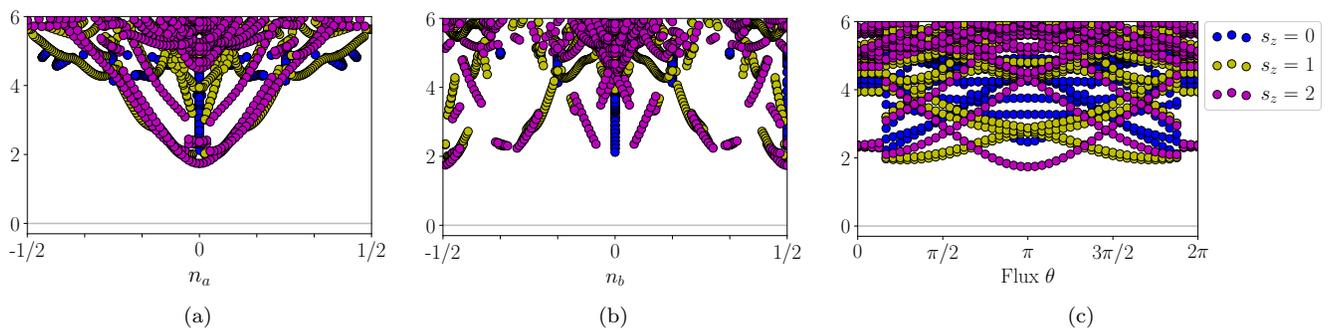

FIG. 10. Transfer matrix spectrum for $(J_2/J, K/J) = (0.1, 0.7)$ near the AFM-AFQ transition, using flux insertion with $\chi = 1000$ and a three-ring unit cell. At quite small flux, around $\pi/6$, there is a first-order transition to some different state, which apparently has low-energy excitations at the $M$ and $\Gamma$ points.

$\theta \to \pi$, which could be a cause for the transition/loss of adiabaticity.

We have not been able to clearly identify the nature of these intermediate-flux states, but we do confirm that they are non-magnetic; furthermore, we observe the emergence of nonzero scalar chirality, though the chirality is small and takes different signs on different plaquettes so that it averages to 0, and thus is likely only a finite bond dimension effect. The main conclusions we can draw are: (1) the flux insertion is $2\pi$-periodic, indicating a lack of spin-fractionalization; and (2) at zero flux there must be a metastable state competing with the ground state. The latter point provides some reason to hope that a spin liquid could still be proximate to the AFM-AFQ transition, and would become the ground state with the addition of some further terms in the Hamiltonian.

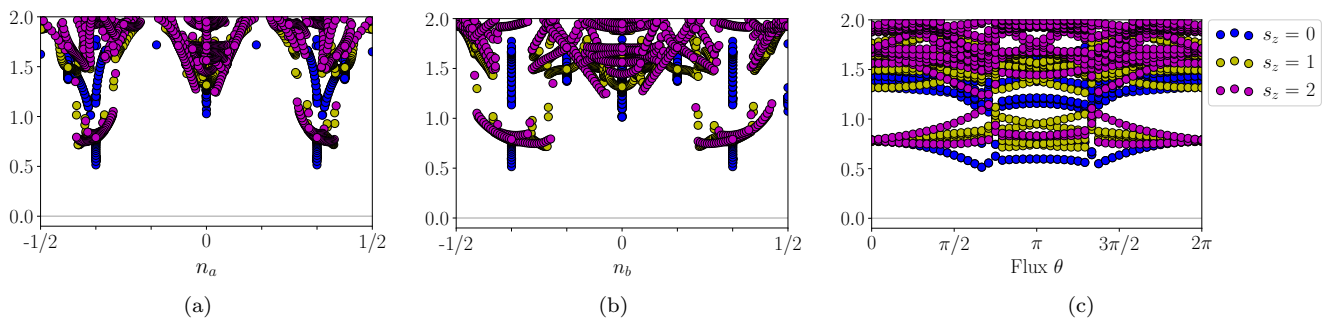

FIG. 11. Transfer matrix spectrum for $(J_2/J, K/J) = (0.1, 0.7)$ near the AFM-AFQ transition, using flux insertion with $\chi = 1000$ and a one-ring unit cell. There is a again a first-order transition, here around $2\pi/3$, possibly caused by an increasing correlation length for $s_z = 0$. The simulation is unstable for a few flux values around $2\pi/3$.

9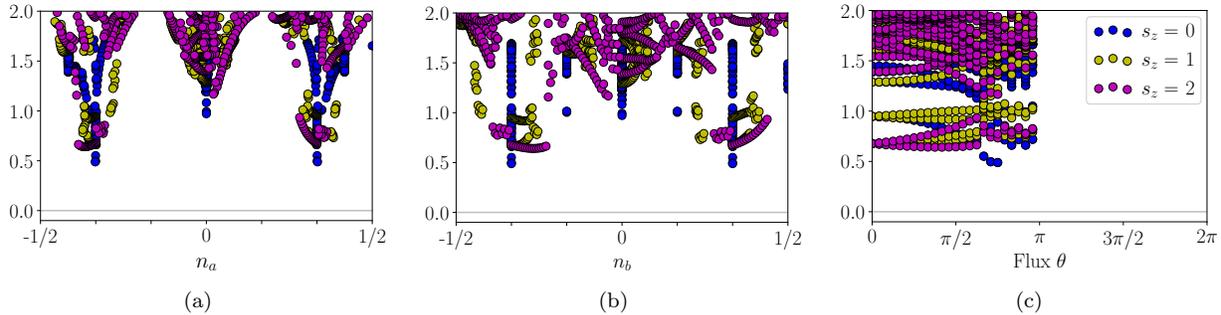

FIG. 12. Transfer matrix spectrum for $(J_2/J, K/J) = (0.1, 0.8)$ in the AFQ phase, using flux insertion with $\chi = 1000$ and a one-ring unit cell. Here the adiabaticity also breaks down around $2\pi/3$, and the DMRG simulations do not converge well from that flux up to $\theta = \pi$.

Finally, we perform flux insertion at a point slightly into the AFQ phase, at $(J_2/J, K/J) = (0.1, 0.8)$, with a one-ring unit cell. As shown in Fig. 12, we see that there is again an apparent first-order transition to some different state. However, the behavior is actually somewhat different. At $K/J = 0.7$, the simulation was unstable for a small range of $\theta$ before converging to the intermediate-flux phase; here, we instead find that the simulation remains unstable between around $2\pi/3$ and $\pi$ and does not converge well to any stable phase.

## IV. SYMMETRY-BREAKING IN THE FERROQUADRUPOLAR PHASE

The cartoon picture of the FQ phase, as described in Section II A of the main text, is that on every site of the lattice, we have the same single-site quadrupolar state $|z\rangle$, or the equivalent along some other direction, $|\hat{n}\rangle$. However, such a state explicitly breaks spin-rotation symmetry, so the true ground state, if SSB does not occur, is something like a superposition of such FQ states with different $\hat{n}$.

In our simulations, there is a further restriction that $\langle S_x^2 \rangle = \langle S_y^2 \rangle$, as we now explain. $S_x = (S^+ + S^-)/2$, while $S_y = (S^+ - S^-)/(2i)$, so that both $S_x^2$ and $S_y^2$ contain the terms $(S^+S^- + S^-S^+)/2$, and they differ only by multiples of $(S^+)^2$ and $(S^-)^2$. We explicitly conserve the quantum number $s_z$, so that the expectation value of any term with a net change in $S_z$ is exactly zero, including both $(S^+)^2$ and $(S^-)^2$. Thus exactly the same terms have nonzero contribution to $\langle S_x^2 \rangle$ and $\langle S_y^2 \rangle$, and we conclude that they must be equal.

Given that we expect SSB at moderate bond dimension (see Appendix A of the main text), this restriction on $\langle S_x^2 \rangle$ and $\langle S_y^2 \rangle$ implies that the only single-site quadrupole that is allowed is $|z\rangle$. On the other hand, we could have a meta-stable state which is approximately an equal superposition of $[|x\rangle$ on every site] and $[|y\rangle$ on every site]. By the same reasoning as in Appendix A, this state would be slightly higher in energy than the one with $|z\rangle$ on every site, but still lower than the symmetry-unbroken state expected as $\chi \to \infty$.

Recall that $|z\rangle$ has $\langle S_z^2 \rangle = 0$ and $\langle S_x^2 \rangle = \langle S_y^2 \rangle = 1$. Then entangled ground state whose cartoon picture is the product state $|z\rangle^{\otimes N}$ will have similar behavior but with reduced quadrupole moment: on each site, $\langle S_z^2 \rangle = x$ for some $x < 2/3$, while $\langle S_x^2 \rangle = \langle S_y^2 \rangle = 1 - x/2$. Correspondingly, the entangled version of $(|x\rangle^{\otimes N} + |y\rangle^{\otimes N})/\sqrt{2}$ would have $\langle S_z^2 \rangle = 1 - x/2$ and $\langle S_x^2 \rangle = \langle S_y^2 \rangle = [(1 - x/2) + x]/2 = 1/2 + x/4$.

We indeed observe both of these states. When we run the simulation for each point in the FQ phase independently, initializing the simulation with a random product state at each parameter point, some converge to the state that is like $|z\rangle^{\otimes N}$, while others converge to the nearly-superposition state. By re-running the simulation at points converging to the nearly-superposition state, initializing with the nearly-$|z\rangle^{\otimes N}$ ground state at a neighboring point, we confirm that the $|z\rangle$-type state is indeed slightly lower in energy throughout the phase, as expected. We can furthermore compare the expectation values of spin fluctuations along $x$, $y$, and $z$ in the ground and the meta-stable states at each parameter point; they are precisely related as given in the preceding paragraph. For example, at $(0.1, -1.0)$, in the $z$-type state we have $(\langle S_x^2 \rangle, \langle S_y^2 \rangle, \langle S_z^2 \rangle) \approx (0.83, 0.83, 0.34)$, whereas in the $(x+y)$-type state we have $\approx (0.58, 0.58, 0.84)$. Both are consistent with $x \approx 0.33 \pm 0.01$.

In Fig. 6 above, the two results in the FQ phase correspond to the $z$-type ground state and $(x+y)$-type metastable state; the former has largest correlation length for $S_z = \pm 1$ and the latter for $S_z = \pm 2$. As bond dimension increases, the energy benefit of breaking symmetry decreases and the $z$-type and $(x+y)$-type states become closer in energy, as shown in Fig. 13. When the bond dimension is sufficiently large, the ground state will be symmetry-preserving, and

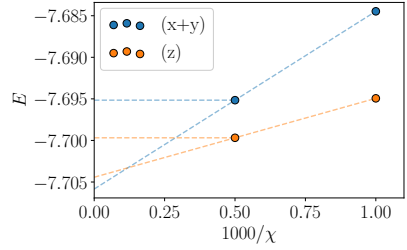

FIG. 13. Energy per site of $z$-type and $(x+y)$-type symmetry-broken FQ states at $(J_2/J, K/J) = (0.1, -1.0)$. The $z$-type state is lower in energy due to the reduced entanglement, but the difference decreases with bond dimension. The wedge between horizontal and diagonal dashed lines shows possible energies in the true $\chi \to \infty$ ground state. We see that the two symmetry-broken states should become degenerate at large enough $\chi$, consistent with the restoration of spin-rotation symmetry.

the $S_z$-resolved correlation lengths will exhibit degeneracy of the longest correlation length for $S_z \in \{-2, -1, 0, 1, 2\}$, corresponding to a spin-2 multiplet as expected for quadrupolar order. With the bond dimensions we consider, we observe this restoration of symmetry at exactly one parameter point, $(J_2/J, K/J) = (0.1, -0.3)$, near the phase boundary.

Additionally, we note that the correlation length is much larger in the $(x+y)$-type states, corresponding to their superposition-like nature. In Fig. 5 of the main text and Fig. 3 above, we use the $(z)$-type state at every parameter point in the FQ phase since it is slightly lower in energy.

## V. ENTANGLEMENT SPECTRA ACROSS THE AFM-AFQ TRANSITION

In Fig. 8 of the main text, we show the evolution of the correlation lengths, for operators carrying different spin and momentum, when crossing the AFM-AFQ transition, and we observe that the transition can be located as the line where the largest correlation lengths for spin-1 and spin-2 become equal. We can also observe the same behavior from the spin- and momentum-resolved entanglement spectra, which we show in Fig. 14. We label the Schmidt values by momentum using a method similar to that discussed in Section III A above, which is explained in Appendix C of the arXiv version of Ref. [5]. Also see the implementation of the function `compute_K` in TenPy [6].

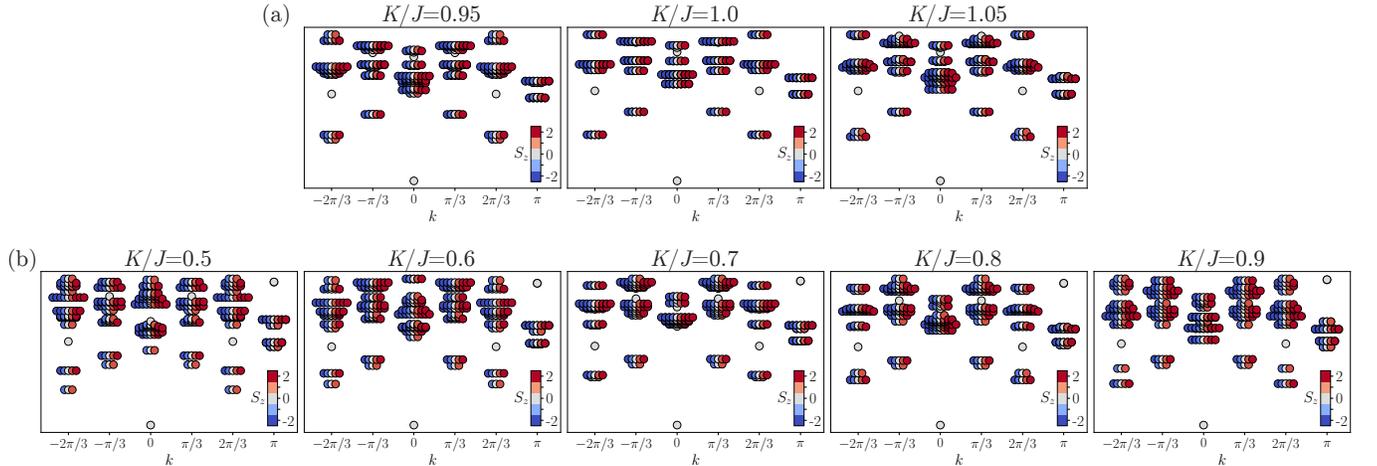

FIG. 14. Spin- and momentum-resolved entanglement spectra across the AFM-AFQ phase transition for (a) $J_2 = 0$, where the transition occurs at $K = J$, and (b) $J_2/J = 0.1$, where the transition occurs at $K \approx 0.7J$. Note that we include a small horizontal offset as a function of the quantum number $S_z$, thus multiplets of 3, 5, or more $S_z$ values in fact share the same momentum; also note that we have cut off the colorbar at $S_z = \pm 2$, although from the horizontal offset we see that higher-spin multiplets are also present.

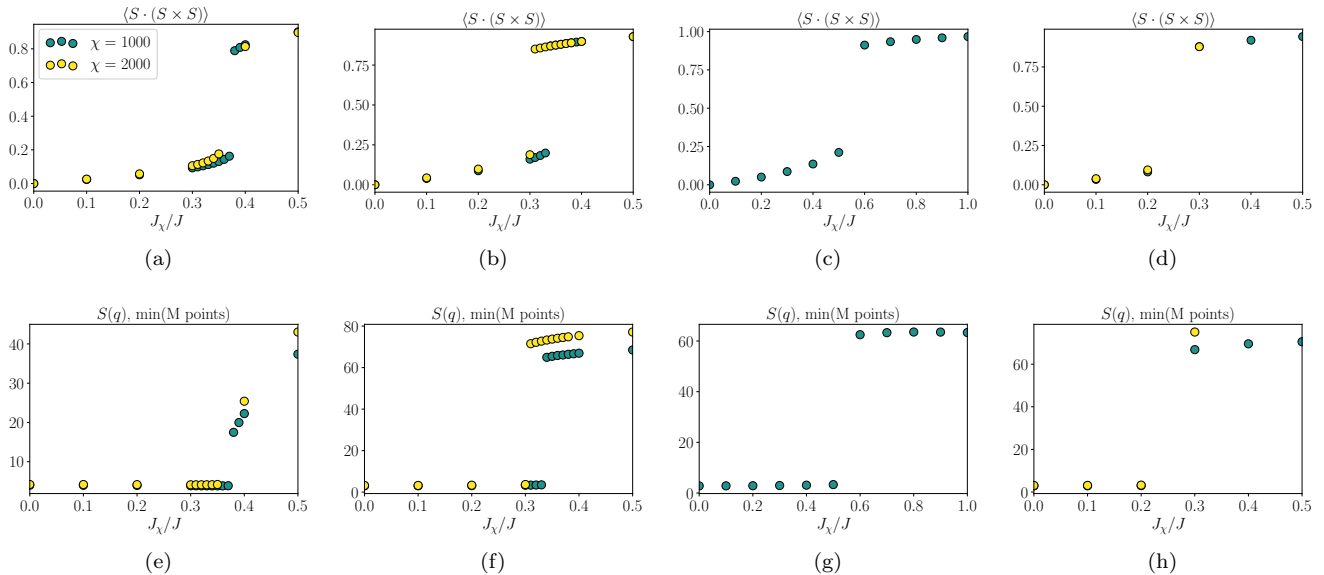

FIG. 15. Scalar chirality $\langle \mathbf{S} \cdot (\mathbf{S} \times \mathbf{S}) \rangle$ [(a)-(d)] and magnetic ordering at the $M$ points [(e)-(h)] as a function of $J_\chi$, for four parameter points: $(J_2/J, K/J) = (0.1, -0.5)$ in the FQ phase [(a),(e)]; $(0.1, 0.8)$ in the AFQ phase [(b),(f)]; $(0, 1)$, the $SU(3)$-symmetric point at the AFM-AFQ transition [(c),(g)]; and $(0.1, 0.7)$, also (approximately) at the AFM-AFQ transition [(d),(h)]. In each case, there is a first-order transition characterized by a sudden increase in both scalar chirality and $M$-point magnetic order, consistent with the tetrahedral magnetic phase.

The spectra shown in Fig. 14 are computed with a one-ring unit cell that disallows three-sublattice symmetry breaking and thus ensures the eigenvalues are organized into spin multiplets. For magnetically ordered states, the lowest spin-1 multiplet should be below the lowest spin-2 multiplet, and for nematic states the opposite should be true. We observe precisely this behavior.

## VI. EFFECT OF SCALAR SPIN CHIRALITY

Further evidence for the lack of a spin liquid phase comes from the addition of a scalar chirality term to the Hamiltonian,

$$H \mapsto H + J_\chi \sum_\Delta \mathbf{S}_i \cdot (\mathbf{S}_j \times \mathbf{S}_k) \tag{10}$$

where the sum is over all plaquettes of the lattice, and $i$, $j$, and $k$ label the three sites of the plaquette. The goal is to check for a time-reversal-symmetric spin liquid that, upon addition of scalar chirality, transitions to a chiral spin liquid (CSL). If that were the case, flux insertion with the chirality included in $H$ would show a quantized spin Hall conductance. Furthermore, the chiral spin liquid would be identifiable by a characteristic entanglement spectrum.

We find no compelling evidence that adding scalar chirality produces a CSL. As shown in Fig. 15, at four different points in the phase diagram, in the FQ and AFQ phases and at the AFM-AFQ transition, the scalar chirality just leads to a first-order transition to the tetrahedral magnetic order; we identify this order by a large scalar chirality and peaks in the spin-structure factor at all three $M$ points. In the figure, there is no indication of a phase transition at any $J_\chi$ below the transition to tetrahedral order, including at $J_\chi = 0$.

We nevertheless consider the possibility that there is also a CSL in this region, which we check by examining the properties mentioned above, the entanglement spectrum and spin pumping with flux insertion. In Fig. 16, we show the spin- and momentum-resolved entanglement spectrum at the $SU(3)$-symmetric point and one other point that would be in a disordered phase between AFM and AFQ if such a phase existed, namely $(J_2/J, K/J) = (0.1, 0.7)$; for each of the two parameter points, we use the largest value of $J_\chi$ below the transition to tetrahedral order—$J_\chi/J = 0.5$ and 0.2, respectively. Both spectra could be plausibly consistent with an $SU(3)_1$ CSL (see Ref. [7], Fig. 9). However, since at the $SU(3)$-symmetric point in particular, the Schmidt values are already organized into multiplets with the correct degeneracy, this is not necessarily a very informative signature of the CSL.

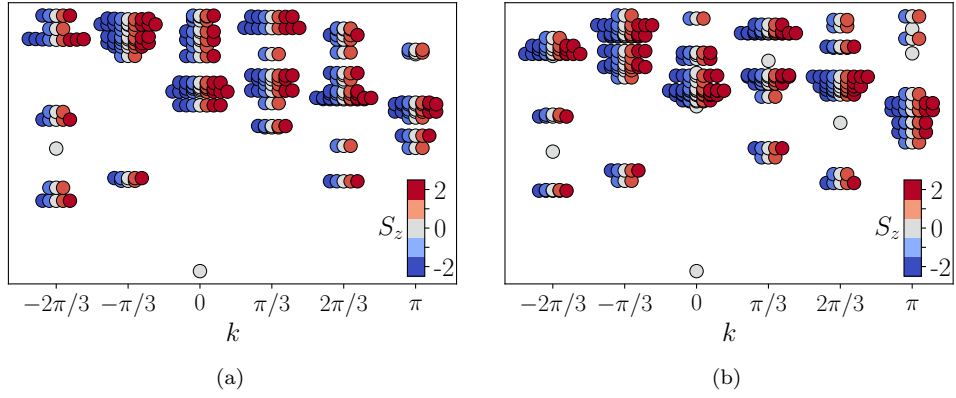

FIG. 16. Spin- and momentum-resolved entanglement spectrum in the ground state with scalar chirality added to the AFM-AFQ transition line, specifically with $(J_2/J, K/J, J_\chi/J) =$ (a) $(0, 1, 0.5)$, and (b) $(0.1, 0.7, 0.2)$. Starting from $k = 0$ and going to the left, both plausibly show the level counting of the $SU(3)_1$ CSL [7], with degeneracies of 1, 8, and 8+1+8.

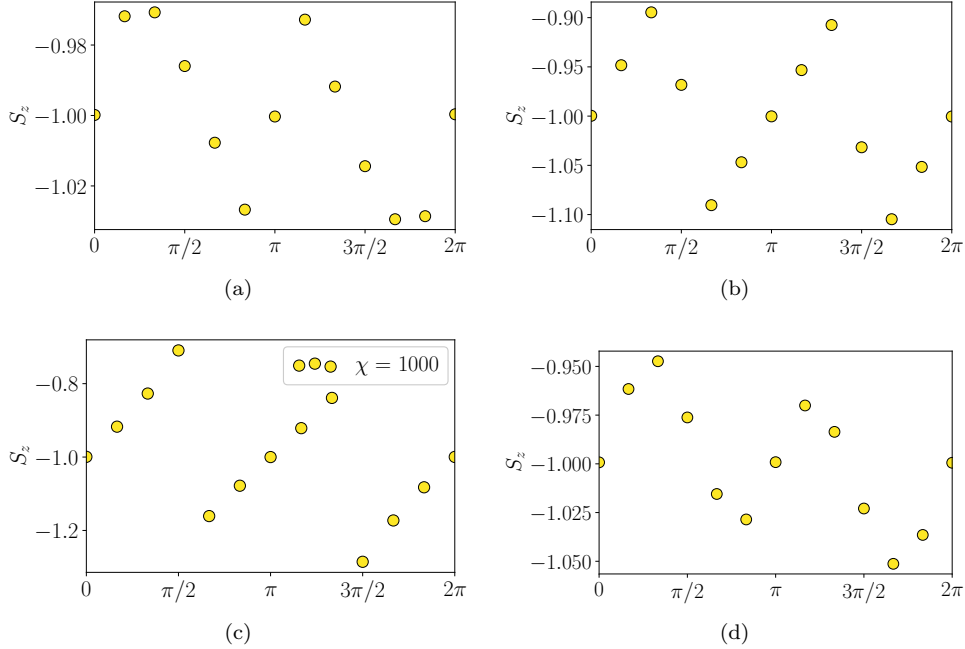

FIG. 17. (a)-(c) Spin pumping at the $SU(3)$-symmetric point with $J_\chi/J$ of (a) 0.3, (b) 0.4, and (c) 0.5. The last one especially looks plausibly consistent with a chiral spin liquid, as explained in the text, but the comparison with the second suggests the spin pumping is not quantized. (d) Spin pumping at $(J_2/J, K/J, J_\chi/J) = (0.1, 0.7, 0.2)$, showing essentially no spin Hall response.

If at these two parameter points, $(J_2/J, K/J, J_\chi/J) = (0, 1, 0.5)$ and $(0.1, 0.7, 0.2)$, we are indeed observing a CSL, there should also be a quantized spin Hall conductance, which manifests in a quantized spin pumping with flux insertion. We perform adiabatic flux insertion, and for each $\theta$ we compute the total ($z$-component of) spin for the left half-cylinder from the entanglement spectrum; note that there is a non-physical constant offset, so only the change with flux insertion is meaningful. The results are shown in Fig. 17.

These data suggest there is no CSL, though there may be an intermediate phase at the $SU(3)$ point with $J_\chi/J$ around 0.4 to 0.5. We see specifically that there is essentially no spin pumping at the $SU(3)$ point with $J_\chi/J = 0.3$ or at $(J_2/J, K/J) = (0.1, 0.7)$ with $J_\chi/J = 0.2$. The two larger values of $J_\chi$ at the $SU(3)$ point do show flux pumping, but with jumps, so the results are hard to interpret. On the one hand, if one remembers the non-physical nature of the overall offset and just takes into account the slope in the segments between the jumps, the $J_\chi/J = 0.5$ data correspond very well to a total spin pumping of exactly 1, which is indeed suggestive of a CSL. On the other hand,

the total spin pumped for $J_\chi/J = 0.4$ is clearly much smaller, so if there is a CSL with a quantized response, its total extent cannot be larger than $J_\chi/J \in (0.4, 0.6)$. (At 0.6, we are already in the tetrahedral magnetic phase.)



\end{document}